\documentclass[journal]{vgtc}
\vgtccategory{Research}
\vgtcpapertype{Technique}

\usepackage[svgnames]{xcolor}%
\usepackage{graphicx}%
\usepackage{booktabs}%
\usepackage{multirow}%
\usepackage{amsmath,amssymb,amsfonts}%
\usepackage{amsthm}%
\usepackage{mathrsfs}%
\usepackage[title]{appendix}%
\usepackage{textcomp}%
\usepackage{manyfoot}%
\usepackage{algorithm}%
\usepackage{algorithmicx}%
\usepackage{algpseudocode}%
\usepackage{listings}%

\usepackage{caption}
\usepackage{subcaption}
\usepackage{tikz}
\usepackage{pgfplots}
\usepackage{array}
\usepackage{amssymb}

\newcommand{\Skip}[1]{}

\newcommand{\revision}[1]{{#1}}

\newcommand{\TODO}[1]{
   \textcolor{red}{\bfseries{TODO: {#1}}}
}

\newcommand{\ToCheck}[1]{{#1}}

\newcommand{\XX}{\textcolor{red}{XX}
}

\begin{document}

\title{RTPD: Penetration Depth calculation using Hardware accelerated Ray-Tracing}

\author{%
  \authororcid{YoungWoo Kim}{0000-0003-3341-1714},
  \authororcid{Sungmin Kwon}{0000-0000-0000-0000}, and
  \authororcid{Duksu Kim}{0000-0002-9075-3983}
}

\authorfooter{
  \item
  	YoungWoo Kim is with Korea University of Technology and Education(KOREATECH).
  	E-mail: aister9@koreatech.ac.kr.
  \item
  	Sungmin Kwon is with Korea University of Technology and Education(KOREATECH).
  	E-mail: 00kwonsm@koreatech.ac.kr.

  \item Duksu Kim is with Korea University of Technology and Education(KOREATECH).
  	E-mail: bluekdct@gmail.com.
}


\abstract{
Penetration depth calculation quantifies the extent of overlap between two objects and is crucial in fields like simulations, the metaverse, and robotics.
Recognizing its significance, efforts have been made to accelerate this computation using parallel computing resources, such as CPUs and GPUs.
Unlike traditional GPU cores, modern GPUs incorporate specialized ray-tracing cores (RT-cores) primarily used for rendering applications.
We introduce a novel algorithm for penetration depth calculation that leverages RT-cores.
Our approach includes a ray-tracing based algorithm for penetration surface extraction and another for calculating Hausdorff distance, optimizing the use of RT-cores.
We tested our method across various generations of RTX GPUs with different benchmark scenes.
The results demonstrated that our algorithm outperformed a state-of-the-art penetration depth calculation method and conventional GPU implementations by up to 37.66 and 5.33 times, respectively. 
These findings demonstrate the efficiency of our RT core-based method and suggest broad applicability for RT-cores in diverse computational tasks.
}
  %

\keywords{GPUs and Multi-core Architectures, Special Purpose Hardware, Geometry-based Techniques, Proximity queries}

\teaser{
  \centering
  \includegraphics[width=\linewidth, alt={Temporal image for teaser.}]{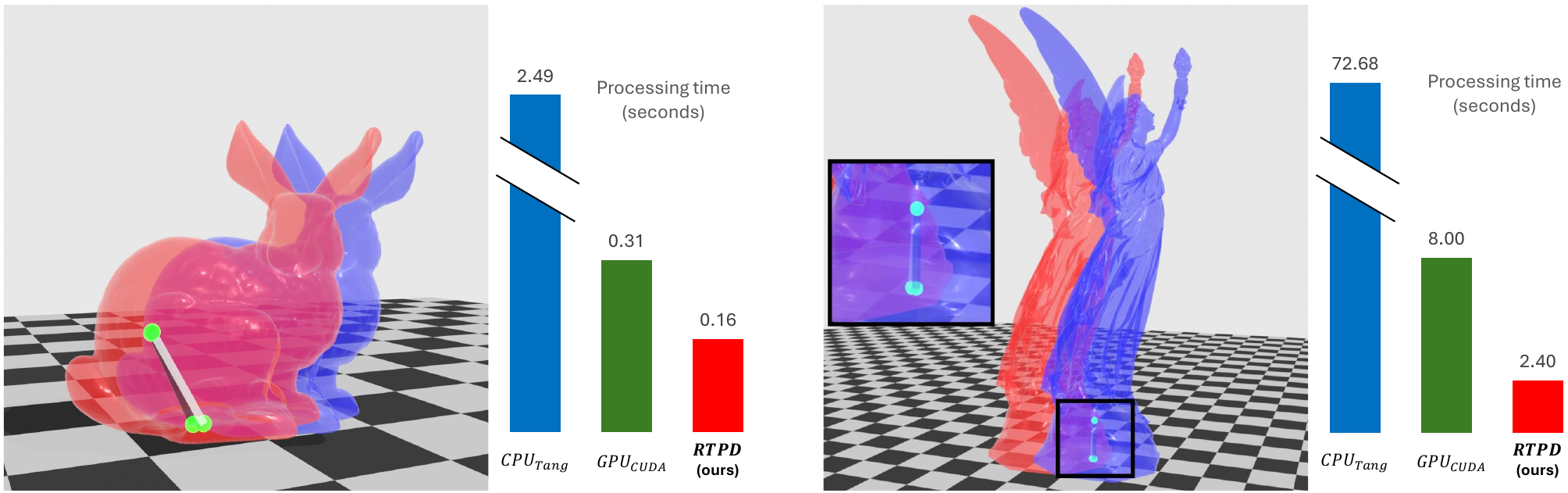}
    \caption{
        Comparison of processing times for penetration depth calculation using the CPU-based implementation based on the state-of-the-art method ($CPU_{Tang}$~\cite{SIG09HIST}), CUDA-based GPU implementation ($GPU_{CUDA}$), and our ray-tracing core-based method ($RTPD$).
        Left: Bunny benchmark. Right: Lucy benchmark.
        The black lines represent the ground truth, while the white lines indicate the results from our method.
        Our method achieved speedups of up to 37.66x and 5.33x compared to $CPU_{Tang}$ and $GPU_{CUDA}$, respectively.
  }
  \label{fig:teaser}
}

\date{February 2025}



\maketitle

\section{Introduction}

Proximity queries, essential for calculating the distance between objects, are fundamental in various fields such as computer graphics, the metaverse, and robotics~\cite{lin2017collision}.
Among these, penetration depth calculation, which measures the extent of overlap between two objects, is widely used in simulations of object interactions and haptic rendering~\cite{zhang2006generalized, laycock2007survey}.

Various methods have been developed for calculating penetration depth.
These methods include Minkowski sum-based algorithms~\cite{dobkin1993computing,je2012polydepth,lee2017penetration} and distance field approaches~\cite{fisher2001fast,sud2006fast}, among others.
Another common technique utilizes the Hausdorff distance between the vertices of the two objects to estimate penetration depth~\cite{SIG09HIST}.
While vertex-based methods provide approximate values and may only offer lower bounds of the penetration depth, they bypass the complexities involved in computing the Minkowski sum, presenting a more straightforward computational approach.

Several parallel processing algorithms have been developed to accelerate proximity query operations, including penetration depth calculations~\cite{kim2002fast,lien2008covering}.
With the advent of General-Purpose GPU (GPGPU) APIs like NVIDIA's Compute Unified Device Architecture (CUDA), the role of graphics processing units (GPUs) has expanded beyond traditional graphics computations to include a variety of general computing tasks.
GPUs are now extensively utilized as acceleration devices in tasks such as proximity computation~\cite{kim2009hpccd,lauterbach2010gproximity,li2011voxelized,kim2013scheduling}, demonstrating their versatility and efficiency in handling computationally intensive operations.

Traditional GPGPU primarily utilizes GPU cores designed for graphics computations.
\revision{However, modern GPUs integrate specialized hardware for ray tracing-based rendering, exemplified by NVIDIA’s RTX (Ray Tracing Texel eXtreme) platform, which introduces dedicated ray-tracing hardware~\cite{turingInDepth}.}
These RTX GPUs are equipped with ray tracing cores (RT cores) optimized for such computations.
Recently, efforts to repurpose RT cores for applications beyond traditional ray tracing have demonstrated high performance and showcased their versatility across various domains~\cite{wald2019rtx,thoman2022multi,zhu2022rtnn,nagarajan2023rt}.

In this work, we propose Ray-Tracing core-based Penetration Depth (RTPD), a novel algorithm for penetration depth computation that leverages RT cores.
Our approach utilizes a Hausdorff distance-based method for measuring penetration depth~\cite{SIG09HIST}.
The RTPD algorithm is structured into two primary components: penetration surface extraction and Hausdorff distance calculation.
To enhance these steps using RT cores, we have developed a ray-tracing based algorithm for penetration point extraction (Sec.~\ref{sec:RT-PIP}) and another for Hausdorff distance calculation (Sec.~\ref{sec:RT-Hausdorff}).
Additionally, we introduce a GPU-based algorithm for generating penetration surfaces (Sec.~\ref{subsec:surfaceGen}), ensuring that RTPD operates entirely on GPU platforms.

To validate the performance of the proposed RTPD algorithm, we implemented it on various generations of RTX GPUs~\cite{thoman2022multi} and tested it across four benchmark scenes featuring models ranging in size from 50K to 12M triangles (Sec.~\ref{sec:result}).
RTPD achieved up to \ToCheck{37.66} times and \ToCheck{5.33} times higher performance than a state-of-the-art CPU algorithm~\cite{zheng2022economic} and a conventional GPU-based implementation, respectively, while maintaining a lower error rate (e.g., less than \ToCheck{6}\%).
These results confirm the effectiveness of the proposed RT core-based penetration depth measurement algorithm and highlight the potential of RT cores for diverse computational applications.

\section{Related Work}

\subsection{Penetration Depth Computation}

The computation of penetration depth often utilizes the Minkowski sum, a well-regarded algorithm documented in Dobkin et al.'s work~\cite{dobkin1993computing}.
This method shows high efficacy for convex shapes, where the simplicity of the objects allows for accurate and computationally efficient penetration depth calculations~\cite{dobkin1993computing,varadhan2004accurate,hachenberger2009exact}.
However, applying this algorithm to concave shapes significantly increases computational complexity.  
As a result, research has focused on developing methods to approximate penetration depth more efficiently for these shapes~\cite{cameron1997enhancing,bergen1999fast,lien2010simple,je2012polydepth}.  

Beyond the Minkowski sum, other methods have been explored, including techniques such as utilizing distance fields or the Hausdorff distance for penetration depth calculations~\cite{fisher2001fast,sud2006fast,SIG09HIST}.

Tang et al.\cite{SIG09HIST} devised an efficient algorithm for calculating the Hausdorff distance between two objects within a given error bound.
They also demonstrated that the proposed algorithm can accelerate penetration depth computation by focusing on the Hausdorff distance in overlapping regions of objects.
Building upon Tang et al.'s method, Zheng et al.\cite{zheng2022economic} improved performance using a BVH-based framework with a four-point strategy.
This method has achieved a performance improvement of up to 20 times compared to Tang et al.'s technique~\cite{SIG09HIST}.
\revision{A common feature of these works, known as the culling-based method, is computing bounds for the Hausdorff distance and reducing the search space.}

\revision{Although culling-based methods have demonstrated significant performance gains, they face challenges in leveraging parallel hardware.  
Updating and sharing bounds require synchronization, which is not well-suited for massively parallel processing architectures such as GPUs.}

\revision{In this work, we propose a GPU-based penetration depth algorithm that specifically accelerates two key processes using RT core technology:  
(1) detecting the overlapping volume and (2) calculating the Hausdorff distance.  
To highlight the effectiveness of our approach, we also implemented a CPU-based penetration depth algorithm based on Tang et al.~\cite{SIG09HIST} and Zheng et al.~\cite{zheng2022economic} for performance comparison.}

\subsection{Ray-Tracing Core-Based Acceleration}

\revision{Recent advancements in GPU technology have led to the integration of dedicated ray-tracing cores (RT cores), enabling hardware-accelerated ray tracing.
These cores optimize intersection checks between rays and objects, allowing for efficient ray-bounding box and ray-triangle intersection tests.
To utilize RT cores, various frameworks such as DXR, OptiX~\cite{parker2010optix}, and Vulkan have been developed.
RT cores primarily accelerate ray intersection tasks by efficiently traversing acceleration hierarchies.}



While the core purpose of ray-tracing cores is to expedite ray tracing, recent studies have explored their application beyond this traditional scope~\cite{wald2019rtx,zhu2022rtnn,thoman2022multi,nagarajan2023rt,meneses2023accelerating,morrical2023attribute}.
Wald et al.~\cite{wald2019rtx} addressed the problem of locating points within tetrahedra using ray-tracing cores.
Zhu et al.~\cite{zhu2022rtnn} introduced a K-Nearest Neighbor (K-NN) algorithm utilizing ray-tracing cores, achieving performance improvements of 2.2 to 65.0 times compared to previous GPU-based nearest neighbor search algorithms.
Thoman et al.~\cite{thoman2022multi} employed RT cores for Room Impulse Response (RIR) simulation.
Nagarajan et al.~\cite{nagarajan2023rt} implemented RT core-based DBSCAN clustering, reporting up to 4 times higher performance enhancement.
Meneses et al.~\cite{meneses2023accelerating} proposed RT core-based Range Minimum Query (RMQ) algorithms, yielding performance up to 2.3 times faster than existing RMQ methods.

\revision{
For collision detection between objects, one of the fundamental proximity queries, researchers have explored ray-tracing approaches even before the introduction of RT-core technology.
Hermann et al.\cite{hermann2008ray} proposed ray-tracing-based collision detection methods for deformable bodies.
Youngjun et al.\cite{kim2010mesh} applied Hermann's idea to medical simulation.
Lehericey et al.\cite{lehericey2015gpu} introduced GPU ray-traced collision detection algorithms for cloth simulation.
Recently, these approaches have been extended to utilize RT cores, as demonstrated by Sui et al.\cite{sui2024hardware}, who proposed discrete and continuous collision detection algorithms using ray-tracing cores.
Unlike these works, which focus on determining when and where collisions occur, our work focuses on calculating penetration depth.
}

In line with these advancements, this study uniquely applies RT-core technology to compute penetration depth, diverging from traditional ray-tracing applications and thereby contributing a novel approach to this field.

\section{Overview}

\revision{In this section, we first explain the foundational concept of Hausdorff distance-based penetration depth algorithms, which are essential for understanding our method (Sec.~\ref{sec:preliminary}).
We then provide a brief overview of our proposed RT-based penetration depth algorithm (Sec.~\ref{subsec:algo_overview}).}

\subsection{Preliminary: Hausdorff Distance-based Penetration Depth}
\label{sec:preliminary}



The Hausdorff distance, as defined in Eq.~\ref{equation:hausdorff_definition}, plays a pivotal role in our approach.
Consider $A$ and $B$ as sets of vertices forming each \revision{object}, and let $d(\cdot, \cdot)$ denote the Euclidean distance between any two vertices.

\begin{equation}
    H(A,B) = \max \left( \max_{a \in A} \min_{b \in B} d(a,b) ,
    \max_{b \in B} \min_{a \in A} d(b,a) \right)
    \label{equation:hausdorff_definition}
\end{equation}

To compute penetration depth using the Hausdorff distance, the process involves several steps.
First, the overlapping volume $V$ between \revision{objects} $A$ and $B$ is computed.
Next, the surfaces of the overlapping volume, $\partial A$ and $\partial B$, contained within each \revision{object}, are extracted.
The final step involves computing the Hausdorff distance $H(\partial A, \partial B)$ between these surfaces.
The resulting distance $H(\partial A, \partial B)$ represents the penetration depth between the two objects.


The brute-force computation of Hausdorff distance has a time complexity of $O(nm)$, where $n$ and $m$ represent the number of vertices in the two objects, as it requires evaluating all vertex pairs.
To reduce this computational burden, Bounding Volume Hierarchy (BVH) is employed, offering rapid localization of target polygons for distance assessment.
Tang et al.~\cite{SIG09HIST} constructed BVHs for two objects, $A$ and $B$.
Their approach begins by computing the Hausdorff distance from $A$ to $B$ (denoted as $h(A,B) = \max_{a \in A} \min_{b \in B} d(a,b)$) through a depth-first traversal of $BVH_A$.
Leveraging the property $h(A', B) \leq h(A, B) \leq h(A, B')$, where $A' \subseteq A$, this step determines the upper bound of the Hausdorff distance, $\overline{h}(A,B)$.
Subsequently, the lower bound $\underline{h}(A,B)$ is determined using $h(B,A)$.
This process yields an approximate Hausdorff distance bound satisfying $\overline{h}(A, B) - \underline{h}(A, B) \leq \epsilon$.
Building upon this method, Zheng et al.~\cite{zheng2022economic} implemented a four-point strategy, sampling four points on triangles (three vertices and one center point) to enable more efficient BVH traversal.
This approach is based on the observation that computing the distance between a triangle and a point is computationally less expensive than computing the distance between two triangles.

\revision{These prior methods, which focused on reducing the search space, achieved significant performance improvements for Hausdorff distance computation.
However, they are not well-suited for parallel processing on GPUs, as they require synchronization for updating and sharing the upper and lower bounds.}

\revision{Departing from previous methods, our approach exploits parallel processing on GPUs while leveraging the intrinsic capabilities of RT cores, with a unique emphasis on the ray-triangle intersection test, which is significantly accelerated by the RT core.
In alignment with Tang et al.~\cite{SIG09HIST}, our method approximates the Hausdorff distance.
However, we place greater emphasis on a ray sampling strategy designed to balance accuracy and performance.
Additionally, while previous methods focused on Hausdorff distance computation and presented penetration depth as an application, we accelerate the entire penetration depth computation process on the GPU.}

\subsection{Overview of the RTPD Algorithm}\label{subsec:algo_overview}
\begin{figure*}[t]
    \centering
    \includegraphics[width=0.8\textwidth]{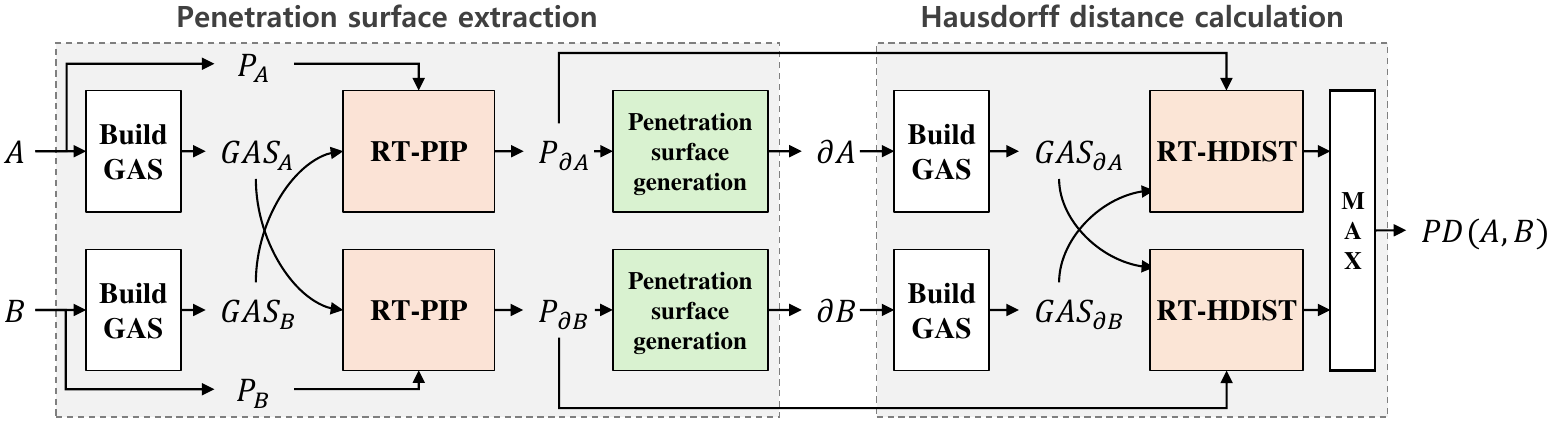}
    \caption{The overview of RT-based penetration depth calculation algorithm overview}
    \label{fig:Overview}
\end{figure*}

Fig.~\ref{fig:Overview} presents an overview of our RTPD algorithm.
It is grounded in the Hausdorff distance-based penetration depth calculation method (Sec.~\ref{sec:preliminary}).
The process consists of two primary phases: penetration surface extraction and Hausdorff distance calculation.
We leverage the RTX platform's capabilities to accelerate both of these steps.

The penetration surface extraction phase focuses on identifying the overlapped region between two objects.
\revision{The penetration surface is defined as a set of polygons from one object, where at least one of its vertices lies within the other object. 
Note that in our work, we focus on triangles rather than general polygons, as they are processed most efficiently on the RTX platform.}
To facilitate this extraction, we introduce a ray-tracing-based \revision{Point-in-Polyhedron} test (RT-PIP), significantly accelerated through the use of RT cores (Sec.~\ref{sec:RT-PIP}).
This test capitalizes on the ray-surface intersection capabilities of the RTX platform.
Initially, a Geometry Acceleration Structure (GAS) is generated for each object, as required by the RTX platform.
The RT-PIP module takes the GAS of one object (e.g., $GAS_{A}$) and the point set of the other object (e.g., $P_{B}$).
It outputs a set of points (e.g., $P_{\partial B}$) representing the penetration region, indicating their location inside the opposing object.
Subsequently, a penetration surface (e.g., $\partial B$) is constructed using this point set (e.g., $P_{\partial B}$) (Sec.~\ref{subsec:surfaceGen}).
The generated penetration surfaces (e.g., $\partial A$ and $\partial B$) are then forwarded to the next step. 

The Hausdorff distance calculation phase utilizes the ray-surface intersection test of the RTX platform (Sec.~\ref{sec:RT-Hausdorff}) to compute the Hausdorff distance between two objects.
We introduce a novel Ray-Tracing-based Hausdorff DISTance algorithm, RT-HDIST.
It begins by generating GAS for the two penetration surfaces, $P_{\partial A}$ and $P_{\partial B}$, derived from the preceding step.
RT-HDIST processes the GAS of a penetration surface (e.g., $GAS_{\partial A}$) alongside the point set of the other penetration surface (e.g., $P_{\partial B}$) to compute the penetration depth between them.
The algorithm operates bidirectionally, considering both directions ($\partial A \to \partial B$ and $\partial B \to \partial A$).
The final penetration depth between the two objects, A and B, is determined by selecting the larger value from these two directional computations.

\section{RT-based Penetration Depth Calculation}

In this section, we provide a detailed explanation of our approaches.

\subsection{RT-based Penetration Point Extraction}\label{sec:RT-PIP}

Identifying whether a vertex $a_i$ from object A is located within the overlapping volume shared by objects A and B involves determining whether $a_i$ is inside object B.
Therefore, we can extract the penetration surface ($\partial A$) of A by identifying all vertices $a_i$ that are inside B and collecting them together.
This method is similarly applied to obtain $\partial B$.
However, this approach requires performing a \revision{point-in-polyhedron} (PIP) test for each vertex in objects A and B, potentially leading to significant computational demands.

We introduce an algorithm, RT-PIP, which leverages the RT-core to accelerate the PIP test.
This is achieved by reinterpreting the PIP test as a series of ray-casting and ray-surface intersection tests.
The underlying principle of this approach is based on the characteristic that if a point is inside an object, a ray cast from this point will intersect the object's surface an odd number of times~\cite{huang1997complexity}.

The initial step of the RT-PIP algorithm involves constructing GASs for each object.
GAS is an essential data structure for performing ray-surface tests with the RT core.
Once the GASs are in place, we begin by casting a ray from a vertex on object $A$ and examining its intersections with object $B$ using the ray-surface intersection capabilities of the RT core.
The RTX platform then returns the count of these intersections.
An odd number of intersections indicates that the vertex from object A is inside object B.

Since the ray’s direction does not influence the PIP test’s outcome, we simply align it along a specific axis, such as (1,0,0).
This method is applied to all vertices $a_i$ in object A to identify those within the overlapping volume ($P_{\partial A}$).
The independence of each vertex’s calculation allows for parallel processing of all vertices, fully utilizing the available RT cores within a GPU.
The same procedure is applied to determine $P_{\partial B}$, identifying all relevant vertices of B in the overlapping volume.

\textbf{Two-way PIP test:}
We found that errors (e.g., false positives) in ray-casting-based PIP tests could arise with RT cores, especially in the region around triangle edges or vertices.
This may be due to the limitations of floating-point precision, as RT cores currently only support floating-point operations.
To address this issue, we employ a two-way ray casting strategy.
In this method, a point is considered to be inside the object (an inner vertex) if it successfully passes the PIP test in both directions, indicating that the number of ray intersections is odd in each case.
This two-way testing strategy enhances the accuracy of our PIP results by mitigating the effects of floating-point precision errors.

\subsection{Penetration Surface Generation on GPU}\label{subsec:surfaceGen}

The RTPD algorithm determines penetration by calculating the Hausdorff distance between penetration surfaces.
Due to the RTX platform's limitation in not supporting the construction of GAS for only parts of an object, it is necessary to create distinct objects for each penetration surface using the extracted penetration points, such as $P_{\partial A}$ and $P_{\partial B}$.
This step requires a new triangle list that includes only the triangles composed of one or more penetration vertices..

To create the penetration triangle list, we traverse the original triangle list once, verifying whether each triangle contains any penetration vertices.
The challenge becomes more complex when constructing a vertex list that includes only penetration vertices, as some vertices are shared by multiple triangles.
This requires assigning a unique vertex index to each penetration vertex to ensure accurate representation and prevent redundancy.

While map data structures offer rapid search and insertion capabilities, making them convenient for creating vertex lists, they present challenges when used in GPU architecture.
Their irregular data access patterns are inefficient for GPUs.
Managing parallel access to maps can also be complex on GPUs, potentially leading to synchronization issues.
Initially, we considered using the CPU to generate the vertex list; however, we found that the overhead of transferring data between the CPU and GPU outweighed the benefits of the RT-core's acceleration capabilities.

\begin{figure}[t]
    \centering
    \includegraphics[width=0.9\linewidth]{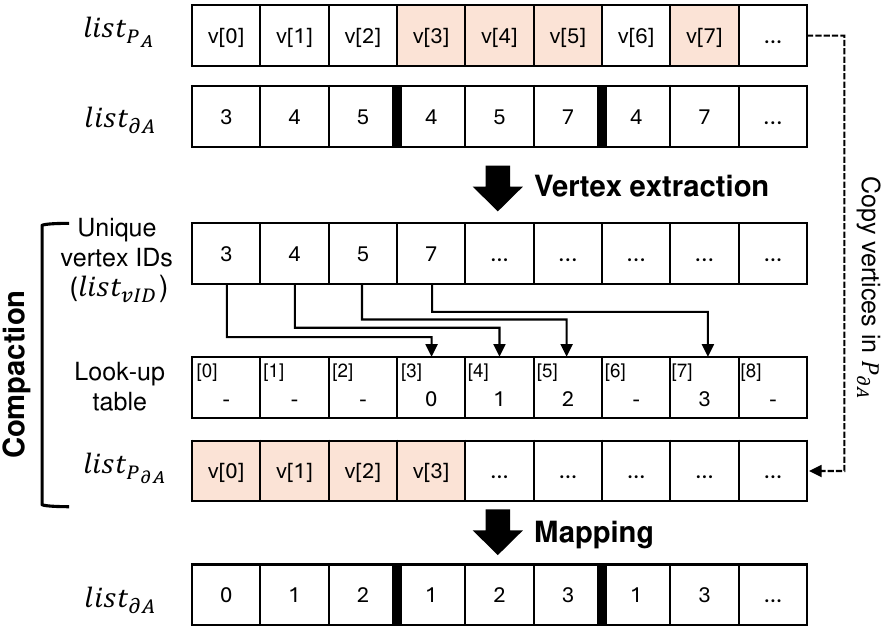}
    \caption{The process of the penetration surface generation on GPU.}
    \label{fig:surfacegen}
\end{figure}

To avoid data communication overhead and fully leverage the GPU's parallel computing capabilities, we developed a CUDA-based algorithm for generating penetration surfaces.
This algorithm consists of three key steps: vertex extraction, compaction, and mapping (Fig.~\ref{fig:surfacegen}).

\textbf{Vertex extraction:}
Each triangle is identified by three vertex indices in the vertex list of the original object ($list_{P_O}$).
This step aims to identify vertices that are part of the penetration surfaces while ensuring no duplicates are present.
The result is a list containing the indices of these unique vertices ($list_{vID}$).
This array is generated by using the vertex index as a key in a reduction operation~\cite{roger2007efficient}.
To enhance GPU performance, the $reduce\_by\_key$ function from NVIDIA's Thrust library is utilized.


\textbf{Compaction:}
The compaction step generates a streamlined vertex list ($list_{P_{\partial O}}$) that includes only the vertices identified in $list_{vID}$ while establishing a lookup table ($lookup$) that maps vertex indices between $list_{P_O}$ and $list_{P_{\partial O}}$.
This process is executed by launching CUDA threads equal to the number of vertices in $list_{vID}$.
Each thread copies a vertex from $list_{P_O}[list_{vID}[t]]$ to $list_{P_{\partial O}}[t]$, where $t$ denotes the thread's ID.
Concurrently, it records the new location of its respective vertex in $list_{P_{\partial O}}$ within the lookup table.
As a result, $lookup[i]$ reflects the revised index within $list_{P_{\partial O}}$ for the $i$-th vertex originally in $list_{P_O}$.


\textbf{Mapping:}
The mapping step updates the vertex indices in $list_{\partial O}$ to correspond with those in $list_{P_{\partial O}}$.
This is accomplished using the lookup table, where each value in $list_{\partial O}$ is replaced by its corresponding index in $list_{P_{\partial O}}$.
In this step, CUDA threads are launched in a quantity equal to the number of triangles in $list_{\partial O}$.

This algorithm is applied to both objects to generate penetration surfaces $\partial A$ and $\partial B$.
Subsequently, the penetration surfaces and penetration points are forwarded to the Hausdorff distance calculation phase.

\subsection{RT-based Hausdorff Distance calculation}\label{sec:RT-Hausdorff}

The penetration depth between objects $A$ and $B$ is equivalent to the Hausdorff distance between their respective penetration surfaces, $\partial A$ and $\partial B$ (i.e., $H(\partial A, \partial B)$).
This distance can be approximated by calculating the distances between the vertices of one object and the surface of the other.
An essential output of the ray-surface intersection test on the RTX platform is the length of the ray from its start point (the vertex) to the intersected surface.
We utilize this ray-surface intersection test capability of the RTX platform to compute the Hausdorff distance, thereby efficiently determining the penetration depth.

Our RT-based Hausdorff distance algorithm, RT-HDIST, calculates both $h(\partial A,\partial B)$ and $h(\partial B,\partial A)$, selecting the larger of the two as the final result.
The process to compute $h(X,Y)$ begins with RT-HDIST acquiring $P_X$ \revision{(the point set of X)} and $GAS_Y$.
The algorithm launches rays from each vertex $p_i$ in $P_X$ toward a set of directions determined by a sampling strategy (Sec.~\ref{subsec:sampling}).
This approach aims to identify the minimum distance from each vertex $p_i$ to object $Y$, denoted as $min_{p_i}$.
RT-HDIST executes this process for each vertex $p_i$ in $P_X$, ultimately determining the maximum value among all $min_{p_i}$ values.
This maximum value represents the calculated directional Hausdorff distance, $h(X,Y)$.
The computation of the Hausdorff distance is enabled by leveraging the ray-surface intersection capabilities of the RTX platform.

\textbf{Ray-length adaptation culling:}
In processing a vertex $p_i$, the goal is to determine the minimum distance from all rays emanating from $p_i$.
Therefore, it is unnecessary to calculate the distance to any surface further than the current minimum distance.
On the RTX platform, the length of a cast ray can be set.
Utilizing this feature, we optimize the algorithm by adapting the ray's length to match the current minimum distance obtained from previously processed rays of $p_i$.
This strategy reduces redundant ray traversal and enhances the efficiency of the RT-HDIST algorithm.

\textbf{Work distribution:}
\revision{Without ray-adaptation culling, each ray-surface intersection test in our algorithm is independent, allowing for concurrent execution.
However, applying ray-adaptation culling requires synchronization among threads handling rays originating from the same vertex, further complicating the process.
Additionally, assigning one thread per ray could lead to an excessive number of threads, increasing overhead associated with thread management.
To address these challenges and improve efficiency, we adopt a work distribution strategy in which a single thread handles all rays associated with a specific vertex, processing them iteratively.
In other words, the basic work unit is a vertex (i.e., a ray source), and our method processes the vertices in parallel.}

\subsection{Ray Sampling Strategy}\label{subsec:sampling}

Although increasing the number of rays improves accuracy, it also increases processing time.
A straightforward approach involves casting rays toward all vertices of the opposing object.
However, this method causes the number of rays to grow exponentially with the penetration surface volume (i.e., $O(|P_{\partial A}| \times |P_{\partial B}|)$).
Additionally, employing this naive strategy results in some rays being directed toward the same or nearly identical directions, leading to redundant ray-surface intersection tests.

One practical approach is the \textit{sphere sampling} method, which entails sampling ray directions directly from the query point.
Sphere sampling is reasonably effective for closely positioned penetration surfaces.
However, it has been observed that as the distance between the penetration surfaces of the two objects increases—especially when the overlapping volume enlarges—the proportion of valid rays, those intersecting the target penetration surface, decreases.
Additionally, the number of rays required to achieve accurate results increases significantly with the volume of overlap, adversely affecting processing performance.

To more accurately target the sampling area, we introduce the \textit{vertex sampling} method.
By selectively sampling vertices from the opposing object, we determine the direction for each ray based on the line extending from the ray origin to the chosen vertex.
We discovered that uniform sampling based on vertex ID is effective in maintaining a balance between accuracy and computational efficiency.
We found that a sampling rate—defined as the ratio of sampled vertices to the total vertices on the penetration surface—of approximately 1\% is generally sufficient to maintain an error ratio below 2\% relative to the ground truth distance. 

Further refining the vertex sampling method, we incorporate the distance information obtained during the PIP test.
When a collision occurs between a ray launched from the query point ($p_i$) and the opposing object, it establishes a specific distance to a point on that object ($d_{pip}(p_i)$).
This distance suggests that the minimum distance between the query point and the object does not exceed $d_{pip}(p_i)$.
Using this information, we optimize the sampling process by focusing only on vertices within this distance, thereby enhancing both the precision and efficiency of our ray sampling strategy.

We have fully integrated the ray sampling processes into CUDA kernels, ensuring that all components of the RT-HDIST algorithm are executed on the GPU.
Furthermore, we have observed that the overhead associated with ray sampling is relatively minor, accounting for less than 0.1\% of the total processing time (e.g., 1-2 ms).

\section{Results and analysis}\label{sec:result}

We implemented our algorithm ($RTPD$) on three different RTX platforms: NVIDIA RTX 2080, NVIDIA RTX 3080, and NVIDIA RTX 4080, which are based on the Turing, Ampere, and Ada Lovelace architectures, respectively (Table~\ref{table:GPUs}).

\begin{table}[t]
\centering
\caption{Main specifications of GPUs used in experiments}
\label{table:GPUs}
\small
\begin{tabular}{|c|c|c|c|}
\hline
GPU                  & RTX2080 & RTX3080 & RTX4080      \\ \hline
Architecture         & Turing  & Ampere  & Ada Lovelace \\ \hline
\# of CUDA   cores   & 2,944    & 8,704    & 9,728        \\ \hline
CUDA core clock             & 1515 Mhz      & 1450 Mhz      & 2205 Mhz           \\ \hline
\# of RT cores        & 46      & 68      & 76           \\ \hline
RT core generation & 1st     & 2nd     & 3rd          \\ \hline
\end{tabular}%
\end{table}



To compare the performance of our method with previous work and conventional GPU implementations, we implemented two alternative methods:


\begin{itemize}
    \item $CPU_{Tang}$ is an implementation of Tang et al.'s method~\cite{SIG09HIST}, with performance further improved by applying Zheng et al.'s upper bound estimation method~\cite{zheng2022economic}.
    However, instead of using the collision detection and hole-filling approaches employed by Tang et al., we implemented a KD-tree-based PIP algorithm using the CGAL library and a Map data structure-based penetration surface generation algorithm, as these demonstrated better performance.
    
    \item $GPU_{cuda}$ is an implementation of the PD algorithm that runs on CUDA cores.
    We adapted an acceleration hierarchy, commonly used in proximity computation~\cite{lauterbach2010gproximity}, to both the PIP and PD stages.
    \revision{We utilized Quantized Bounding Volume Hierarchies (QBVH4) for their proven efficiency on GPUs~\cite{wald2019rtx}, building them directly on the GPU.
    Specifically, we sort the triangles using Morton codes when reading the object and then build the tree in a bottom-up manner.}
    For PD computation, we designed the system so that each thread handles a query point, maximizing the GPU’s parallel computing capabilities.
    For penetration surface generation, we employed our GPU-based algorithm presented in Sec.~\ref{subsec:surfaceGen}, enabling $GPU_{cuda}$ to execute entirely on the GPU.

\end{itemize}

All test systems were equipped with identical CPUs (Intel i5-14600K) and 32GB of system memory.
For implementing our method and the alternatives, we used CUDA 12.3, Optix SDK 7.4, and the NVIDIA Thrust Library.

\textbf{Benchmarks: }
We established four distinct benchmark scenes using well-known models ranging in size from 50K to 12M triangles.
Each model was preprocessed using a hole-filling algorithm to ensure the objects were closed.
For each benchmark, we positioned two identical objects differently to achieve varying overlap ratios from 0.1 to 0.9.
Table~\ref{table_bench} provides detailed information about these benchmarks.
We established the ground truth for penetration depth through a brute-force computation of all vertex pairs, carefully separating overlapping volumes into distinct regions to account for the potential complexity of objects with multiple overlapping areas.

\begin{table*}[]
\centering
\caption{Benchmark scenes used in the experiments, showcasing various overlap ratios for different models.
The black lines represent the ground truth penetration depths, while the white lines indicate the results obtained using our method.}
\label{table_bench}
\small
\begin{tabular}{@{}c|c|@{}m{2.7cm}@{ }m{2.7cm}@{ }m{2.7cm}@{ }m{2.7cm}@{ }m{2.7cm}@{ }m{2.7cm}@{ }m{2.7cm}@{ }m{2.7cm}@{}}
\hline
\multirow{2}{*}{Benchmark} & \# of Vtx. (\# of Tri.) & \multicolumn{5}{c}{Scene (overlap ratio)} \\  \cline{3-7}
    & per object              
    & \multicolumn{1}{c}{0.1}
    & \multicolumn{1}{c}{0.3}
    & \multicolumn{1}{c}{0.5}
    & \multicolumn{1}{c}{0.7}
    & \multicolumn{1}{c}{0.9}
    \\ \hline
{$Bunny$}                    & 0.26M (0.53M)           
    & \includegraphics[width=\linewidth]{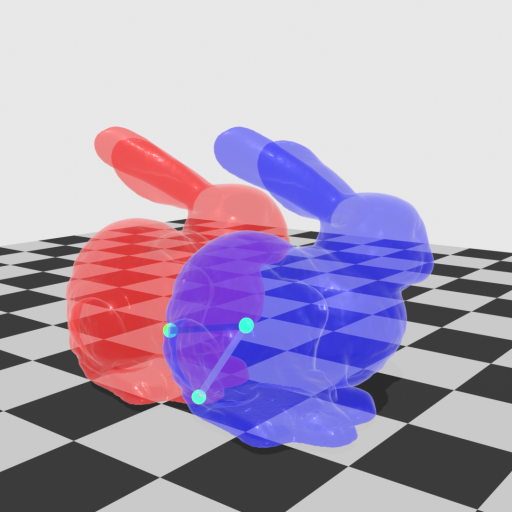}   
    & \includegraphics[width=\linewidth]{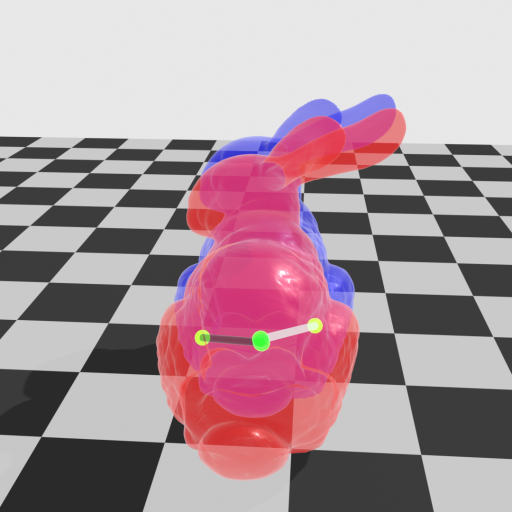}       
    & \includegraphics[width=\linewidth]{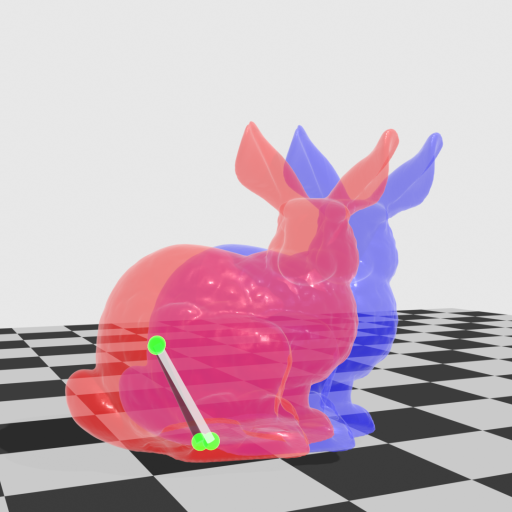}        
    & \includegraphics[width=\linewidth]{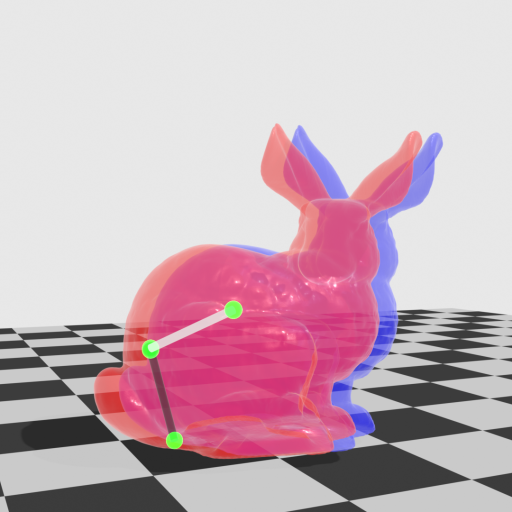}        
    & \includegraphics[width=\linewidth]{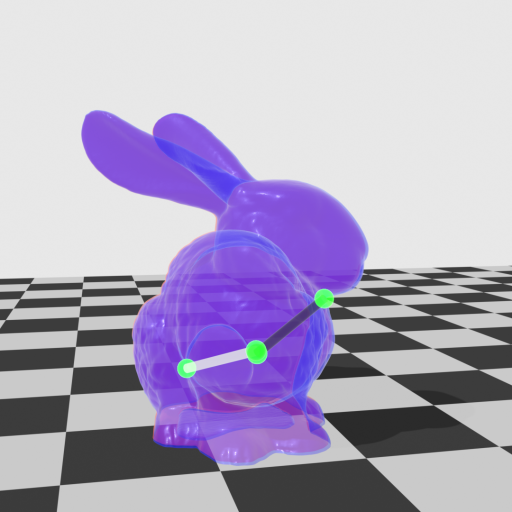}       
    \\ \hline
{$David$}                     & 0.44M (0.88M)            
    & \includegraphics[width=\linewidth]{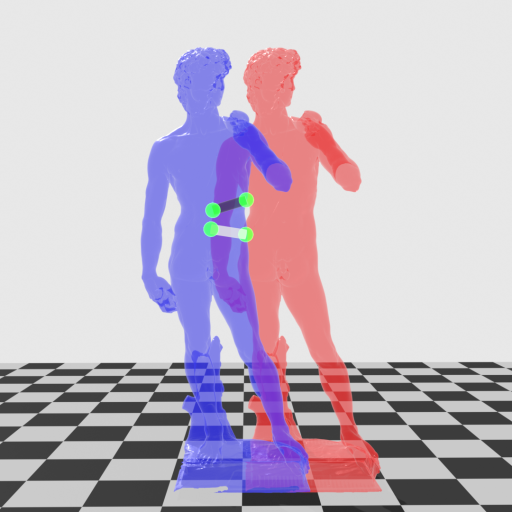}   
    & \includegraphics[width=\linewidth]{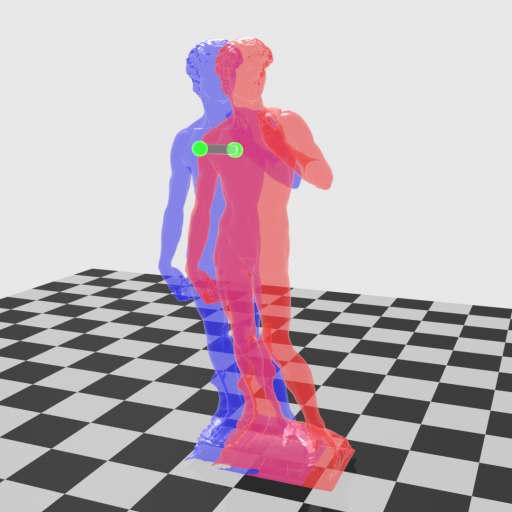}       
    & \includegraphics[width=\linewidth]{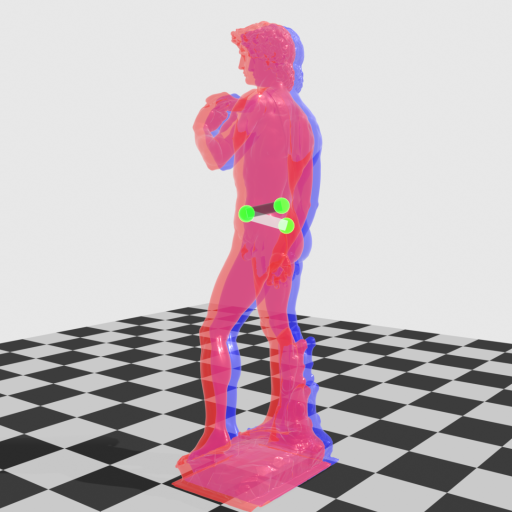}        
    & \includegraphics[width=\linewidth]{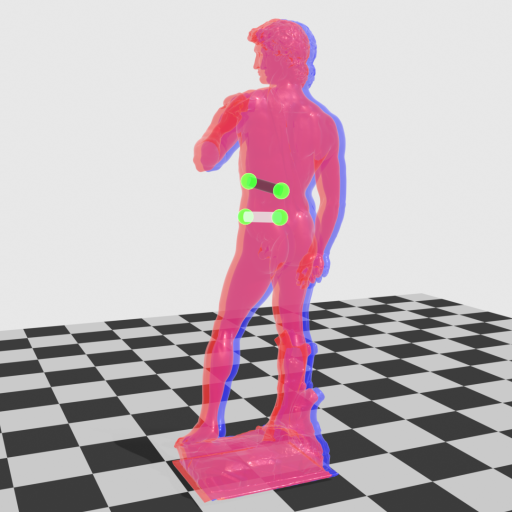}        
    & \includegraphics[width=\linewidth]{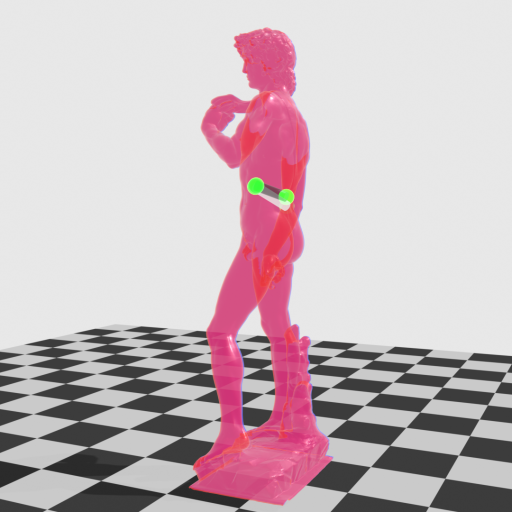}       
    \\ \hline
{$St.~Matthew$}                 & 5.96M (11.92M)          
    & \includegraphics[width=\linewidth]{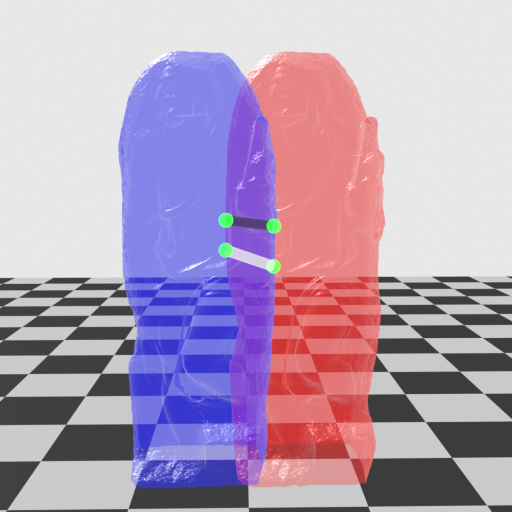}   
    & \includegraphics[width=\linewidth]{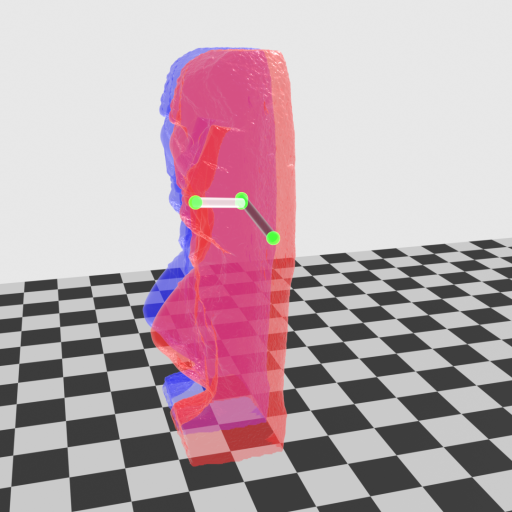}       
    & \includegraphics[width=\linewidth]{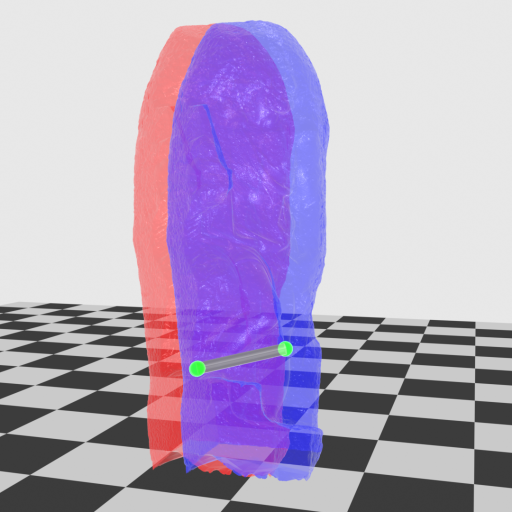}        
    & \includegraphics[width=\linewidth]{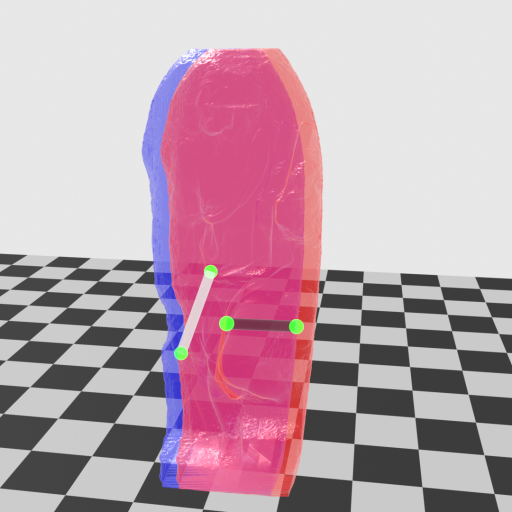}        
    & \includegraphics[width=\linewidth]{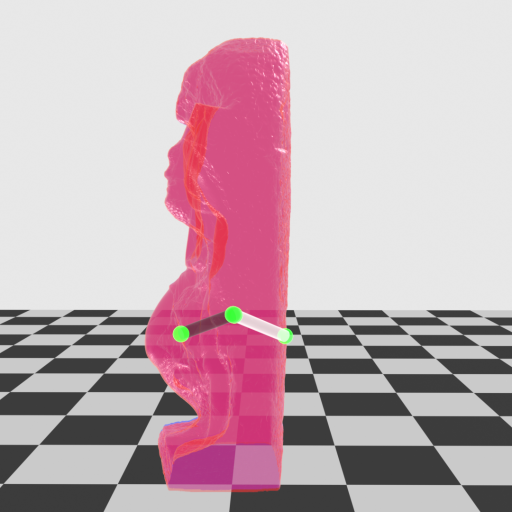}       
    \\ \hline
{$Lucy$}                     & 6.42M (12.85M)          
    & \includegraphics[width=\linewidth]{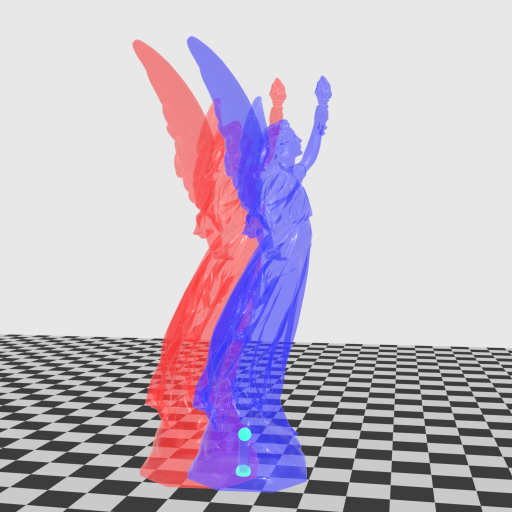}   
    & \includegraphics[width=\linewidth]{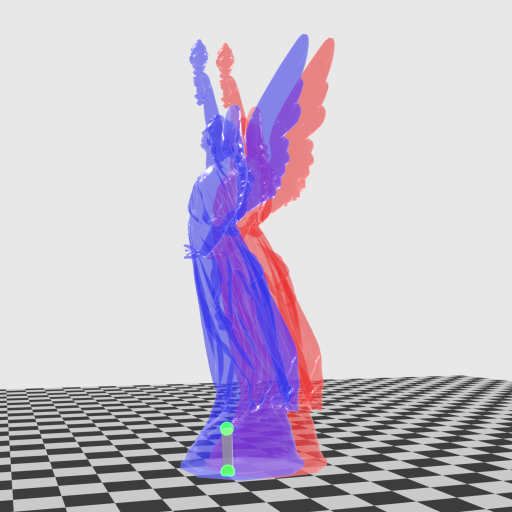}       
    & \includegraphics[width=\linewidth]{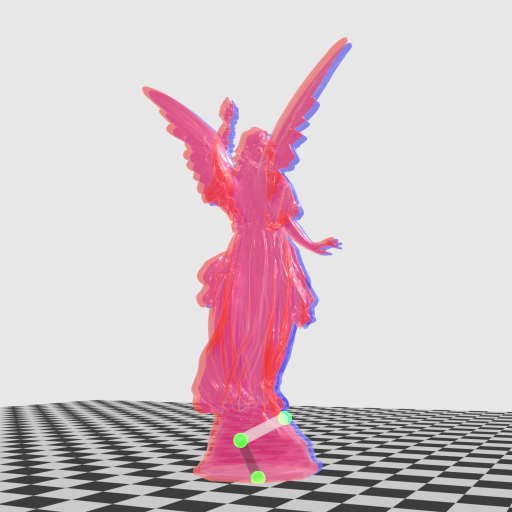}        
    & \includegraphics[width=\linewidth]{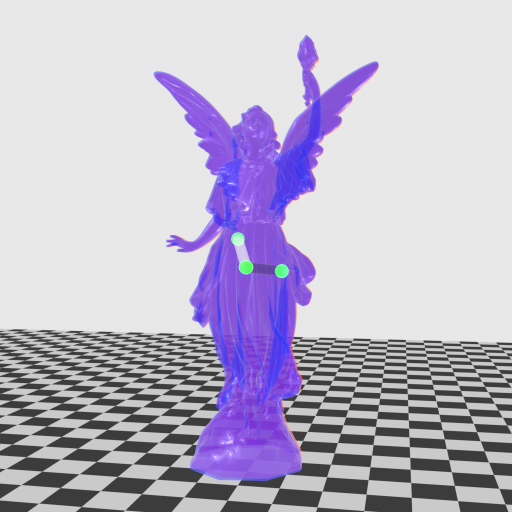}        
    & \includegraphics[width=\linewidth]{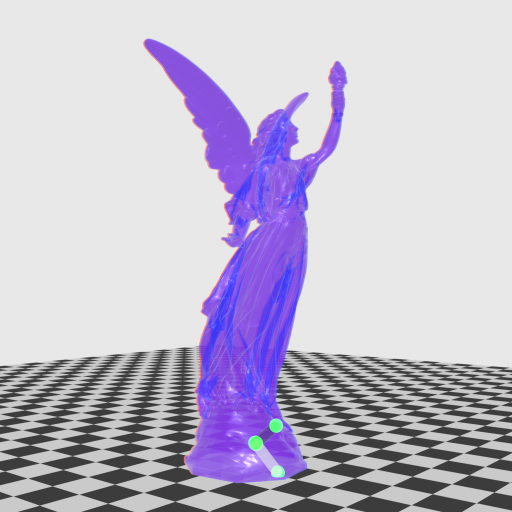}       
    \\ \hline
\end{tabular}%
\end{table*}




\Skip{
\begin{figure*}[htb!]
    \centering
    \includegraphics[width=0.9\textwidth]{Image/result_image.pdf}
    \caption{Visualization of penetration depth result using $baseline$ and $RTPD$.}
    \label{fig:image_result}
\end{figure*}
}

\subsection{Results}

\begin{table*}[]
\centering
\small
\caption{This table shows the computation times (in milliseconds) for penetration depth using three different algorithms across four benchmarks.}
\label{table:result}
\begin{tabular}{cccccccccccc}
\hline
\multicolumn{2}{|c|}{Benchmark} &
  \multicolumn{5}{c|}{$Bunny$} &
  \multicolumn{5}{c|}{$David$} \\ \hline
\multicolumn{2}{|c|}{Overlap ratio} &
  \multicolumn{1}{c|}{0.1} &
  \multicolumn{1}{c|}{0.3} &
  \multicolumn{1}{c|}{0.5} &
  \multicolumn{1}{c|}{0.7} &
  \multicolumn{1}{c|}{0.9} &
  \multicolumn{1}{c|}{0.1} &
  \multicolumn{1}{c|}{0.3} &
  \multicolumn{1}{c|}{0.5} &
  \multicolumn{1}{c|}{0.7} &
  \multicolumn{1}{c|}{0.9} \\ \hline
\multicolumn{2}{|c|}{$CPU_{Tang}$} &
  \multicolumn{1}{c|}{1,349.97} &
  \multicolumn{1}{c|}{1,971.65} &
  \multicolumn{1}{c|}{2,494.34} &
  \multicolumn{1}{c|}{2,923.93} &
  \multicolumn{1}{c|}{3,397.98} &
  \multicolumn{1}{c|}{2,530.55} &
  \multicolumn{1}{c|}{3,822.21} &
  \multicolumn{1}{c|}{5,280.25} &
  \multicolumn{1}{c|}{5,944.63} &
  \multicolumn{1}{c|}{6,525.23} \\ \hline
\multicolumn{1}{|c|}{\multirow{2}{*}{\begin{tabular}[c]{@{}c@{}}RTX\\ 2080\end{tabular}}} &
  \multicolumn{1}{c|}{$GPU_{cuda}$} &
  \multicolumn{1}{c|}{335.21} &
  \multicolumn{1}{c|}{540.96} &
  \multicolumn{1}{c|}{611.64} &
  \multicolumn{1}{c|}{760.13} &
  \multicolumn{1}{c|}{854.56} &
  \multicolumn{1}{c|}{487.52} &
  \multicolumn{1}{c|}{877.72} &
  \multicolumn{1}{c|}{1,382.58} &
  \multicolumn{1}{c|}{1,515.79} &
  \multicolumn{1}{c|}{1,780.15} \\ \cmidrule{2-12} 
\multicolumn{1}{|c|}{} &
  \multicolumn{1}{c|}{$RTPD$} &
  \multicolumn{1}{c|}{81.73} &
  \multicolumn{1}{c|}{138.68} &
  \multicolumn{1}{c|}{179.45} &
  \multicolumn{1}{c|}{204.48} &
  \multicolumn{1}{c|}{247.02} &
  \multicolumn{1}{c|}{140.53} &
  \multicolumn{1}{c|}{243.06} &
  \multicolumn{1}{c|}{394.40} &
  \multicolumn{1}{c|}{457.15} &
  \multicolumn{1}{c|}{529.75} \\ \hline
\multicolumn{1}{|c|}{\multirow{2}{*}{\begin{tabular}[c]{@{}c@{}}RTX\\ 3080\end{tabular}}} &
  \multicolumn{1}{c|}{$GPU_{cuda}$} &
  \multicolumn{1}{c|}{170.72} &
  \multicolumn{1}{c|}{279.66} &
  \multicolumn{1}{c|}{305.27} &
  \multicolumn{1}{c|}{376.56} &
  \multicolumn{1}{c|}{432.13} &
  \multicolumn{1}{c|}{272.39} &
  \multicolumn{1}{c|}{384.01} &
  \multicolumn{1}{c|}{658.94} &
  \multicolumn{1}{c|}{711.40} &
  \multicolumn{1}{c|}{833.11} \\ \cmidrule{2-12} 
\multicolumn{1}{|c|}{} &
  \multicolumn{1}{c|}{$RTPD$} &
  \multicolumn{1}{c|}{87.55} &
  \multicolumn{1}{c|}{128.08} &
  \multicolumn{1}{c|}{165.00} &
  \multicolumn{1}{c|}{177.69} &
  \multicolumn{1}{c|}{213.25} &
  \multicolumn{1}{c|}{130.08} &
  \multicolumn{1}{c|}{208.98} &
  \multicolumn{1}{c|}{310.78} &
  \multicolumn{1}{c|}{375.77} &
  \multicolumn{1}{c|}{420.57} \\ \hline
\multicolumn{1}{|c|}{\multirow{2}{*}{\begin{tabular}[c]{@{}c@{}}RTX\\ 4080\end{tabular}}} &
  \multicolumn{1}{c|}{$GPU_{cuda}$} &
  \multicolumn{1}{c|}{113.49} &
  \multicolumn{1}{c|}{148.10} &
  \multicolumn{1}{c|}{157.05} &
  \multicolumn{1}{c|}{175.48} &
  \multicolumn{1}{c|}{191.35} &
  \multicolumn{1}{c|}{133.69} &
  \multicolumn{1}{c|}{200.61} &
  \multicolumn{1}{c|}{291.89} &
  \multicolumn{1}{c|}{309.29} &
  \multicolumn{1}{c|}{347.82} \\ \cmidrule{2-12} 
\multicolumn{1}{|c|}{} &
  \multicolumn{1}{c|}{$RTPD$} &
  \multicolumn{1}{c|}{69.60} &
  \multicolumn{1}{c|}{95.83} &
  \multicolumn{1}{c|}{121.98} &
  \multicolumn{1}{c|}{138.94} &
  \multicolumn{1}{c|}{163.73} &
  \multicolumn{1}{c|}{108.91} &
  \multicolumn{1}{c|}{172.60} &
  \multicolumn{1}{c|}{246.23} &
  \multicolumn{1}{c|}{288.23} &
  \multicolumn{1}{c|}{323.81} \\ \hline
\multicolumn{1}{l}{} &
  \multicolumn{1}{l}{} &
  \multicolumn{1}{l}{} &
  \multicolumn{1}{l}{} &
  \multicolumn{1}{l}{} &
  \multicolumn{1}{l}{} &
  \multicolumn{1}{l}{} &
  \multicolumn{1}{l}{} &
  \multicolumn{1}{l}{} &
  \multicolumn{1}{l}{} &
  \multicolumn{1}{l}{} &
  \multicolumn{1}{l}{} \\ \hline
\multicolumn{2}{|c|}{Benchmark} &
  \multicolumn{5}{c|}{$stmatthew$} &
  \multicolumn{5}{c|}{$Lucy$} \\ \hline
\multicolumn{2}{|c|}{Overlap ratio} &
  \multicolumn{1}{c|}{0.1} &
  \multicolumn{1}{c|}{0.3} &
  \multicolumn{1}{c|}{0.5} &
  \multicolumn{1}{c|}{0.7} &
  \multicolumn{1}{c|}{0.9} &
  \multicolumn{1}{c|}{0.1} &
  \multicolumn{1}{c|}{0.3} &
  \multicolumn{1}{c|}{0.5} &
  \multicolumn{1}{c|}{0.7} &
  \multicolumn{1}{c|}{0.9} \\ \hline
\multicolumn{2}{|c|}{$CPU_{Tang}$} &
  \multicolumn{1}{c|}{20,814.80} &
  \multicolumn{1}{c|}{36,688.20} &
  \multicolumn{1}{c|}{46,477.20} &
  \multicolumn{1}{c|}{57,628.90} &
  \multicolumn{1}{c|}{64,832.50} &
  \multicolumn{1}{c|}{72,682.00} &
  \multicolumn{1}{c|}{79,554.70} &
  \multicolumn{1}{c|}{102,890.00} &
  \multicolumn{1}{c|}{130,233.00} &
  \multicolumn{1}{c|}{147,261.00} \\ \hline
\multicolumn{1}{|c|}{\multirow{2}{*}{\begin{tabular}[c]{@{}c@{}}RTX\\ 2080\end{tabular}}} &
  \multicolumn{1}{c|}{$GPU_{cuda}$} &
  \multicolumn{1}{c|}{19,816.20} &
  \multicolumn{1}{c|}{36,432.50} &
  \multicolumn{1}{c|}{57,616.00} &
  \multicolumn{1}{c|}{73,974.10} &
  \multicolumn{1}{c|}{67,363.10} &
  \multicolumn{1}{c|}{16,738.50} &
  \multicolumn{1}{c|}{25,704.30} &
  \multicolumn{1}{c|}{40,311.40} &
  \multicolumn{1}{c|}{65,276.80} &
  \multicolumn{1}{c|}{73,633.70} \\ \cmidrule{2-12} 
\multicolumn{1}{|c|}{} &
  \multicolumn{1}{c|}{$RTPD$} &
  \multicolumn{1}{c|}{4,429.51} &
  \multicolumn{1}{c|}{9,193.99} &
  \multicolumn{1}{c|}{13,054.40} &
  \multicolumn{1}{c|}{18,727.20} &
  \multicolumn{1}{c|}{25,086.60} &
  \multicolumn{1}{c|}{3,138.68} &
  \multicolumn{1}{c|}{7,055.98} &
  \multicolumn{1}{c|}{13,021.90} &
  \multicolumn{1}{c|}{19,532.80} &
  \multicolumn{1}{c|}{24,348.70} \\ \hline
\multicolumn{1}{|c|}{\multirow{2}{*}{\begin{tabular}[c]{@{}c@{}}RTX\\ 3080\end{tabular}}} &
  \multicolumn{1}{c|}{$GPU_{cuda}$} &
  \multicolumn{1}{c|}{9,602.12} &
  \multicolumn{1}{c|}{18,445.30} &
  \multicolumn{1}{c|}{29,214.30} &
  \multicolumn{1}{c|}{38,252.20} &
  \multicolumn{1}{c|}{34,912.90} &
  \multicolumn{1}{c|}{8,002.66} &
  \multicolumn{1}{c|}{12,110.20} &
  \multicolumn{1}{c|}{19,306.90} &
  \multicolumn{1}{c|}{32,417.80} &
  \multicolumn{1}{c|}{36,425.80} \\ \cmidrule{2-12} 
\multicolumn{1}{|c|}{} &
  \multicolumn{1}{c|}{$RTPD$} &
  \multicolumn{1}{c|}{3,372.21} &
  \multicolumn{1}{c|}{6,538.83} &
  \multicolumn{1}{c|}{8,699.78} &
  \multicolumn{1}{c|}{11,846.90} &
  \multicolumn{1}{c|}{15,362.40} &
  \multicolumn{1}{c|}{2,401.75} &
  \multicolumn{1}{c|}{5,175.98} &
  \multicolumn{1}{c|}{8,911.14} &
  \multicolumn{1}{c|}{12,655.40} &
  \multicolumn{1}{c|}{14,554.30} \\ \hline
\multicolumn{1}{|c|}{\multirow{2}{*}{\begin{tabular}[c]{@{}c@{}}RTX\\ 4080\end{tabular}}} &
  \multicolumn{1}{c|}{$GPU_{cuda}$} &
  \multicolumn{1}{c|}{3,485.11} &
  \multicolumn{1}{c|}{5,928.27} &
  \multicolumn{1}{c|}{8,771.09} &
  \multicolumn{1}{c|}{10,682.00} &
  \multicolumn{1}{c|}{10,088.30} &
  \multicolumn{1}{c|}{3,055.14} &
  \multicolumn{1}{c|}{4,389.57} &
  \multicolumn{1}{c|}{6,637.78} &
  \multicolumn{1}{c|}{9,800.50} &
  \multicolumn{1}{c|}{10,311.70} \\ \cmidrule{2-12} 
\multicolumn{1}{|c|}{} &
  \multicolumn{1}{c|}{$RTPD$} &
  \multicolumn{1}{c|}{2,497.75} &
  \multicolumn{1}{c|}{4,786.06} &
  \multicolumn{1}{c|}{6,276.15} &
  \multicolumn{1}{c|}{8,434.39} &
  \multicolumn{1}{c|}{10,804.20} &
  \multicolumn{1}{c|}{1,930.04} &
  \multicolumn{1}{c|}{3,965.84} &
  \multicolumn{1}{c|}{6,678.99} &
  \multicolumn{1}{c|}{9,183.59} &
  \multicolumn{1}{c|}{10,637.80} \\ \hline
\end{tabular}%
\end{table*}

Table~\ref{table:result} displays the processing times for three different algorithms, including ours and two alternatives.
For $RTPD$, we utilized vertex sampling rates of 1.5\% for $Bunny$ and $David$, and 0.45\% for $St.~Matthew$ and $Lucy$, respectively.

\begin{figure}[t]
    \centering
    \includegraphics[width=0.95\columnwidth]{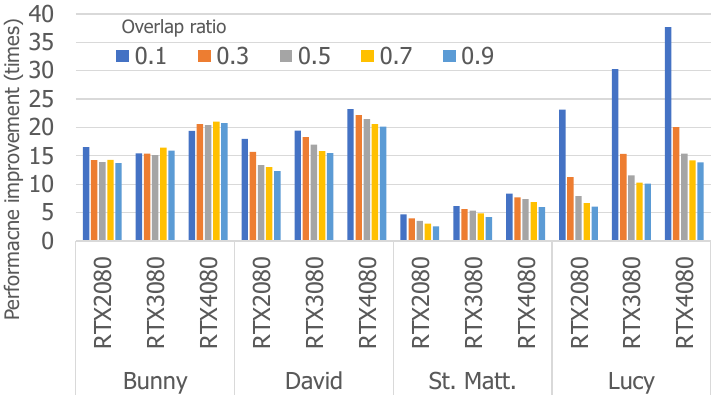}
    \caption{Performance improvement of $RTPD$ over $CPU_{Tang}$ across various benchmarks with differing overlap ratios on different RTX GPUs.}
    \label{fig:perf_over_cpu}
\end{figure}

Compared to $CPU_{Tang}$, our method demonstrated up to 37.66 times (13.89 times on average) higher performance (Fig.~\ref{fig:perf_over_cpu}).
The extent of performance improvement varies across different scenes, with $RTPD$ achieving more significant enhancements on more powerful GPUs equipped with a greater number of RT cores.
Specifically, $RTPD$ achieved average performance improvements of 9.77, 12.38, and 16.27 times on RTX 2080, 3080, and 4080, respectively.
These results confirm that our method scales effectively with the number of RT cores.

\begin{figure}[t]
    \centering
    \includegraphics[width=0.95\columnwidth]{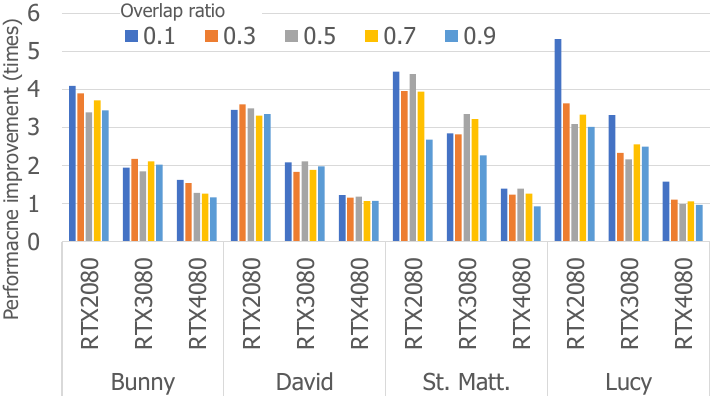}
    \caption{Performance improvement of $RTPD$ over $GPU_{cuda}$ across various benchmarks with differing overlap ratios on different RTX GPUs.}
    \label{fig:perf_over_gpu}
\end{figure}

$GPU_{cuda}$, efficiently utilizing CUDA cores, achieved up to 23.79 times (7.68 times on average) higher performance compared to $CPU_{Tang}$.
On RTX GPUs, $RTPD$ demonstrated up to 5.33 times (2.43 times on average) better performance than $GPU_{cuda}$ (Fig.~\ref{fig:perf_over_gpu}).


We found that the performance gap between $RTPD$ and $GPU_{cuda}$ is less significant on the RTX 4080 compared to other GPUs.
Specifically, $RTPD$ achieved average performance improvements of 3.50, 2.33, and 1.18 times on the RTX 2080, 3080, and 4080, respectively.
This narrowing gap can be attributed to the significant increase in the number of CUDA cores and their clock speeds in the RTX 4080, which contrasts with the more modest increment in RT cores (Table~\ref{table:GPUs}).
Nevertheless, $RTPD$ consistently outperforms $GPU_{cuda}$ across all tested GPU platforms.
Additionally, it is important to note that CUDA cores and RT cores are independent processing resources, allowing for their simultaneous utilization.

\begin{table}[t]
\centering
\small
\caption{
The error rate over the ground truth.
}
\label{table:error}
\begin{tabular}{|c|c|c|c|c|c|}
\hline
Overlap   ratio & 0.1    & 0.3    & 0.5    & 0.7    & 0.9    \\ \hline
$Bunny$         & 0.61\% & 0.45\% & 1.59\% & 1.24\% & 1.88\% \\ \hline
$David$         & 0.38\% & 1.54\% & 1.42\% & 1.84\% & 2.25\% \\ \hline
$St.~matthew$     & 0.30\% & 4.68\% & 0.65\% & 5.56\% & 2.22\% \\ \hline
$Lucy$          & 0.09\% & 0.68\% & 0.92\% & 1.00\% & 3.61\% \\ \hline
\end{tabular}
\end{table}
Table~\ref{table:error} presents the error rates of $RTPD$ in relation to the ground truth.
The error rate varies depending on several factors including the sampling rate and the seed used for sampling initiation (e.g., starting vertex ID).
Additionally, model complexity (e.g., the number of triangles) and the size of the overlap volume also influence accuracy.
For instance, with smaller objects such as $Bunny$ and $David$, a very low sampling rate results in an insufficient number of rays for achieving accurate results, even though it may substantially improve performance compared to alternative methods.
In contrast, for larger models like $St.~Matthew$ and $Lucy$, a low sampling rate often suffices for maintaining reasonable accuracy (e.g., less than 2\%) because the polygons are more densely packed to depict finer details of the objects.
In our experiments, the sampling rates were set at 1.5\% for $Bunny$ and $David$, and 0.45\% for $St.~Matthew$ and $Lucy$ to balance accuracy with computational efficiency.
A thorough analysis of the impact of sampling rates on accuracy and performance is provided in Sec.~\ref{subsec:sampling_rate}.

\subsection{Performance Analysis}

To investigate the impact of RT cores on the performance enhancement of our proposed method compared to other algorithms, we measured the processing times of three key steps across these algorithms. 

Table~\ref{table:breakdown} presents an analysis of results on an RTX 3080 GPU for benchmarks with an overlap ratio of 0.5, offering detailed insights into efficiency improvements at each computational stage.  
For this section, we focus our analysis on this specific experiment, as the performance trends observed here align with those in other benchmark scene setups.
\begin{table}[t]
\centering
\caption{
This table compares the processing times (in milliseconds) of three algorithms across three major computational steps: penetration point extraction (\revision{PPE}), penetration surface generation (PSG), and Hausdorff distance calculation (HDIST), executed on an RTX 3080 for an overlap ratio of 0.5.
The values in parentheses indicate the performance improvement factor over the $CPU_{Tang}$ algorithm.}
\label{table:breakdown}
\resizebox{\columnwidth}{!}{%
\begin{tabular}{|c|c|c|c|c|}
\hline
 Bench.                     & Algo.         & \revision{PPE}              & PSG             & HDIST           \\ \hline \hline
\multirow{3}{*}{$Bunny$}    & $CPU_{Tang}$ & 863.31      & 215.44     & 1,467.59    \\ \cmidrule{2-5} 
                            & $GPU_{cuda}$  & 89.75 (9.62x)    & 7.58 (28.44x)   & 201.92 (7.27x)  \\ \cmidrule{2-5} 
                            & $RTPD$        & 44.48 (19.41x)   & 8.58 (25.12x)   & 111.84 (13.12x) \\ \hline \hline
\multirow{3}{*}{$David$}    & $CPU_{Tang}$ & 1,972.02     & 437.39     & 3,005.79    \\ \cmidrule{2-5} 
                            & $GPU_{cuda}$  & 175.18 (11.26x)  & 9.14 (47.86x)   & 465.38 (6.46x)  \\ \cmidrule{2-5} 
                            & $RTPD$        & 51.25 (38.48x)   & 11 (39.78x)     & 248.37 (12.10x)  \\ \hline \hline
\multirow{3}{*}{$St. Matthew$} & $CPU_{Tang}$ & 27,381.30     & 6,384.10     & 47,187.30    \\ \cmidrule{2-5} 
                            & $GPU_{cuda}$  & 2,342.24 (11.69x) & 82.39 (77.49x)  & 26,665.10 (1.77x) \\ \cmidrule{2-5} 
                            & $RTPD$        & 547.45 (50.02x)  & 100.56 (63.49x) & 8,051.55 (5.86x) \\ \hline \hline
\multirow{3}{*}{$Lucy$}     & $CPU_{Tang}$ & 48,308.40     & 6,313.96    & 48,249.40    \\ \cmidrule{2-5} 
                            & $GPU_{cuda}$  & 3,315.28 (14.57x) & 83.64 (75.49x)  & 15,900.30 (3.03x) \\ \cmidrule{2-5} 
                            & $RTPD$        & 493.49 (97.89x)  & 101.21 (62.39x) & 8,316.22 (5.80x)  \\ \hline
\end{tabular}%
}
\end{table}

In the penetration point extraction step (i.e., \revision{PPE} in Table~\ref{table:breakdown}), the RT-based algorithm (\revision{RT-PPE}) achieved up to 97.89 times (51.45 times on average) higher performance compared to the CPU-based algorithm.
When contrasted with the CUDA implementation ($GPU_{cuda}$), RT-PPE showed up to a 6.42 times (4.11 times on average) increase in performance.
Unlike $GPU_{cuda}$, which traverses the BVH using general-purpose CUDA cores, RT-PPE benefits significantly from the accelerated BVH traversal enabled by specialized RT-core hardware, leading to these notable performance gains.

For the penetration surface generation step (PSG in Table~\ref{table:breakdown}), the GPU-based algorithm exhibits 57.32 times higher performance on average compared to the CPU method.
Since $RTPD$ employs the same algorithm for this step, the processing times should be identical. However, it takes slightly longer, likely due to the overhead associated with context switching between RT cores and CUDA cores.
Nevertheless, the PSG step constitutes a minor portion of the overall processing time, making the context-switching overhead negligible in terms of overall performance.

For the Hausdorff distance calculation (HDIST in Table~\ref{table:breakdown}), the RT-based algorithm (RT-HDIST) achieved up to 13.12 times (9.22 times on average) higher performance compared to $CPU_{Tang}$, and up to 3.31 times (2.23 times on average) better performance relative to $GPU_{cuda}$.
Although the performance gain over $GPU_{cuda}$ is less dramatic compared to the PPE step—largely due to the extensive ray processing required—RT-HDIST consistently demonstrates a performance advantage over traditional CUDA-based methods.

\begin{figure}[t]
    \centering
    \includegraphics[width=0.9\columnwidth]{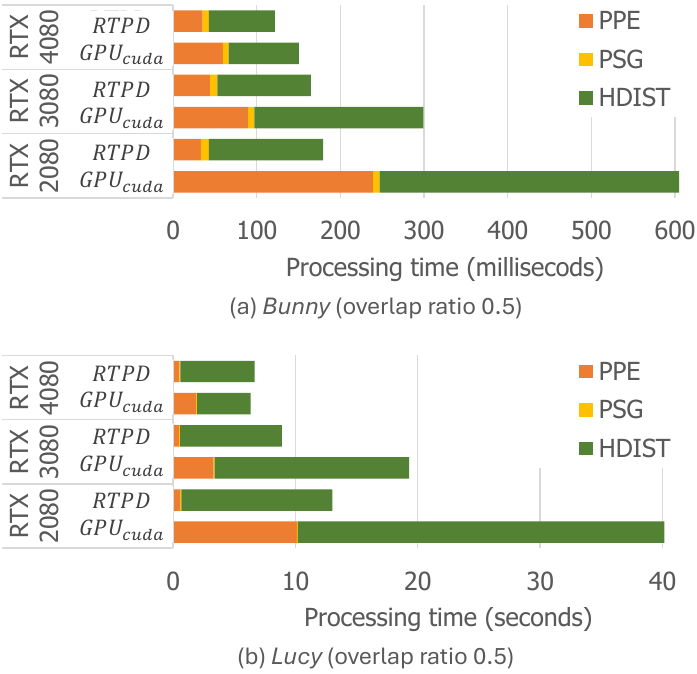}
    \caption{The graphs illustrate the processing times for $GPU_{cuda}$ and $RTPD$ across three major computational stages—\revision{PPE}, PSG, and HDIST—on different RTX GPUs.}
    \label{fig:analysis}
\end{figure}

The stacked column charts in Fig.~\ref{fig:analysis} illustrate the processing times for each computational step of two GPU-based algorithms, $GPU_{cuda}$ and $RTPD$, across three different RTX GPUs.
\Skip{
\begin{figure}[]
    \centering
    \begin{subfigure}{\columnwidth}
        \includegraphics[width=\textwidth]{Image/graphs/analysis_bunny.pdf}
        \caption{$Bunny$ (overlap ratio 0.5)}
        \vspace{0.2cm}
        \label{subfig:analysis_bunny}
    \end{subfigure}
    \begin{subfigure}{\columnwidth}
        \includegraphics[width=\textwidth]{Image/graphs/analysis_lucy.pdf}
        \caption{$Lucy$ (overlap ratio 0.5)}
        \label{subfig:analysis_lucy}
    \end{subfigure}
    \caption{These graphs shows}
    \label{fig:anaysis}
\end{figure}
}

Fig.~\ref{fig:analysis}-(a) and \ref{fig:analysis}-(b) display results for scenes with small and large-scale objects, respectively, highlighting the performance of each algorithm at the PPE, PSG, and HDIST stages.
The PSG step, across all evaluated scenarios, accounts for a relatively minor portion of the total computational effort, and its impact is particularly negligible in large-scale scenes.

Among the three stages, HDIST is consistently the most time-intensive.
On the RTX 2080—the least powerful GPU in the test—HDIST consumes over 60\% and 70\% of the total processing time for $GPU_{cuda}$ and $RTPD$, respectively, in scenarios involving small-scale objects (e.g., $Bunny$).
The dominance of the HDIST stage increases with scene complexity; for instance, in large-scale scenarios such as $Lucy$, it accounts for more than 90\% of $RTPD$'s processing time.
Both $GPU_{cuda}$ and $RTPD$ continue to improve the performance of the HDIST step with more powerful GPUs.
However, compared to $GPU_{cuda}$, $RTPD$ exhibits less scalability with newer GPU generations, likely due to the more substantial advancements in CUDA cores compared to RT cores across generations (Table~\ref{table:GPUs}).
Despite this, $RTPD$ consistently outperforms $GPU_{cuda}$ in all cases, underscoring its efficacy in leveraging RT cores for penetration depth calculations.

The PPE step represents the second most significant computational overhead.
For $GPU_{cuda}$, this step accounts for about 20–40\% of the total processing time.
In contrast, our RT-PPE demonstrates much higher efficiency than the CUDA-based PPE, contributing only about 5–30\% of the total computational load for $RTPD$.
Notably, the overhead of the PPE step becomes almost negligible in $RTPD$ for large-scale scenes, as shown in Fig.~\ref{fig:analysis}-(b).
One notable observation is that the processing time for the PPE step remains relatively consistent across the three different GPUs in small-scale benchmarks such as $Bunny$ and $David$.
This consistency arises because a significant portion of the time in RT-PPE (more than 90\%) is spent generating the GAS, and the number of rays—two per vertex—is relatively small, thus not fully utilizing the extensive RT-core capabilities.
However, in large-scale benchmarks, the performance of RT-PPE is notably scalable.
For instance, the PPE time for $Lucy$ is 10.12, 3.31, and 1.90 milliseconds on RTX 2080, 3080, and 4080, respectively (Fig.~\ref{fig:analysis}-(b)).

\textbf{Effect of ray-length adaptation culling:}
To evaluate the effect of ray-length adaptation culling in RT-HDIST (Sec.~\ref{sec:RT-Hausdorff}), we measured the processing time of the HDIST step both with and without this method.
The results are shown in Table~\ref{table_culling}. 
We observed that ray-length adaptation culling reduced processing time by up to 70\%.
This method consistently improved performance across all benchmarks and overlap ratios, achieving an average performance increase of 2.21 times compared to when culling was not applied.
\begin{table}[t]
\centering
\caption{Processing time (in milliseconds) of the HDIST step, executed on an RTX 3080, with (+) and without (-) ray-length adaptation culling.}
\label{table_culling}
\resizebox{\columnwidth}{!}{%
\begin{tabular}{|c|c|ccccc|}
\hline
\multirow{2}{*}{Bench.} &
  \multirow{2}{*}{Cull.} &
  \multicolumn{5}{c|}{Overlap ratio} \\ \cmidrule{3-7} 
 &
   &
  \multicolumn{1}{c|}{0.1} &
  \multicolumn{1}{c|}{0.3} &
  \multicolumn{1}{c|}{0.5} &
  \multicolumn{1}{c|}{0.7} &
  0.9 \\ \hline \hline
\multirow{2}{*}{$Bunny$} &
  - &
  \multicolumn{1}{c|}{56.44} &
  \multicolumn{1}{c|}{108.16} &
  \multicolumn{1}{c|}{162.96} &
  \multicolumn{1}{c|}{221.95} &
  279.10 \\ \cmidrule{2-7} 
 &
  + &
  \multicolumn{1}{c|}{31.30} &
  \multicolumn{1}{c|}{75.62} &
  \multicolumn{1}{c|}{111.84} &
  \multicolumn{1}{c|}{125.26} &
  159.38 \\ \hline \hline
\multirow{2}{*}{$David$} &
  - &
  \multicolumn{1}{c|}{99.02} &
  \multicolumn{1}{c|}{239.98} &
  \multicolumn{1}{c|}{481.20} &
  \multicolumn{1}{c|}{607.05} &
  708.32 \\ \cmidrule{2-7} 
 &
  + &
  \multicolumn{1}{c|}{65.27} &
  \multicolumn{1}{c|}{146.43} &
  \multicolumn{1}{c|}{248.37} &
  \multicolumn{1}{c|}{312.89} &
  357.26 \\ \hline \hline
\multirow{2}{*}{$Stmat.$} &
  - &
  \multicolumn{1}{c|}{7,729.89} &
  \multicolumn{1}{c|}{14,835.60} &
  \multicolumn{1}{c|}{21,350.70} &
  \multicolumn{1}{c|}{28,872.10} &
  37,823.30 \\ \cmidrule{2-7} 
 &
  + &
  \multicolumn{1}{c|}{2,753.27} &
  \multicolumn{1}{c|}{5,905.73} &
  \multicolumn{1}{c|}{8,051.55} &
  \multicolumn{1}{c|}{11,189.00} &
  14,697.60 \\ \hline \hline
\multirow{2}{*}{$Lucy$} &
  - &
  \multicolumn{1}{c|}{6,091.47} &
  \multicolumn{1}{c|}{10,984.20} &
  \multicolumn{1}{c|}{22,796.30} &
  \multicolumn{1}{c|}{32,652.00} &
  38,153.10 \\ \cmidrule{2-7} 
 &
  + &
  \multicolumn{1}{c|}{1,831.17} &
  \multicolumn{1}{c|}{4,600.55} &
  \multicolumn{1}{c|}{8,316.22} &
  \multicolumn{1}{c|}{12,045.70} &
  13,935.90 \\ \hline
\end{tabular}
}%
\end{table}

\subsection{Impact of sampling rate}\label{subsec:sampling_rate}
Our method involves sampling a given number of vertices from the overlap surface to compute the Hausdorff distance, significantly impacting the accuracy of the results.
To evaluate how variations in the sampling rate affect accuracy, we conducted a thorough analysis of both accuracy and processing time in relation to different sampling rates.

\begin{figure}[t]
    \centering
    \includegraphics[width=\columnwidth]{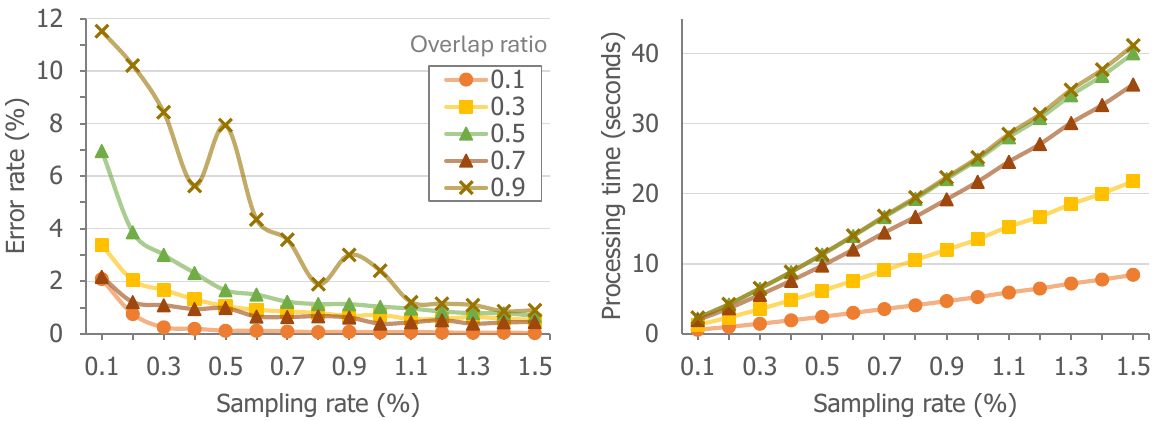}
    \caption{The graphs display changes in error rates as the sampling rate increases in the RT-HDIST algorithm for the $Lucy$ benchmark across various overlap ratios.}
    \label{fig:sampling_rate}
\end{figure}

Fig.~\ref{fig:sampling_rate} presents the results for the $Lucy$ benchmark, with each line representing outcomes for different overlap ratios.
Acknowledging the inherent randomness in the sampling process, which can cause fluctuations in error rates even at the same sampling rate, we conducted ten trials for each sampling rate using distinct sampling seeds and calculated the average to ensure a comprehensive assessment of performance consistency.


As anticipated, the error rate decreases as the sampling rate increases, and the processing time tends to rise almost linearly with the sampling rate.
It is notable that larger overlap volumes require higher sampling rates to ensure accuracy.
Additionally, our results indicate that the robustness to sampling seed variability diminishes with larger overlap volumes, as illustrated for an overlap ratio of 0.9 in Fig.~\ref{fig:sampling_rate}.
Nevertheless, for smaller overlap volumes, which are common in applications like physics-based simulations and haptic rendering, our method demonstrates stable performance even with lower sampling rates and random sampling seeds.
These results demonstrate the effectiveness and applicability of our $RTPD$ algorithm across a variety of applications.

\subsection{Comparison of sampling strategies}

To evaluate the efficiency of the vertex sampling method we employed, we compared the error rate changes in our method as the sampling rate increased against two alternatives.
The first alternative is sphere sampling, which samples ray directions from the surface of a sphere enclosing the query point.
The second alternative is an Axis-Aligned Bounding Box (AABB)-based ray sampling method, designed to more precisely target the sampling area.
This method selects points within the AABB surrounding the penetration surface of the opposing object and determines the direction of each ray based on the line extending from the ray origin to the sampled point.
For both alternative methods, we applied ray-length adaptation culling for a fair comparison.

\begin{figure}[t]
    \centering
    \includegraphics[width=0.8\columnwidth]{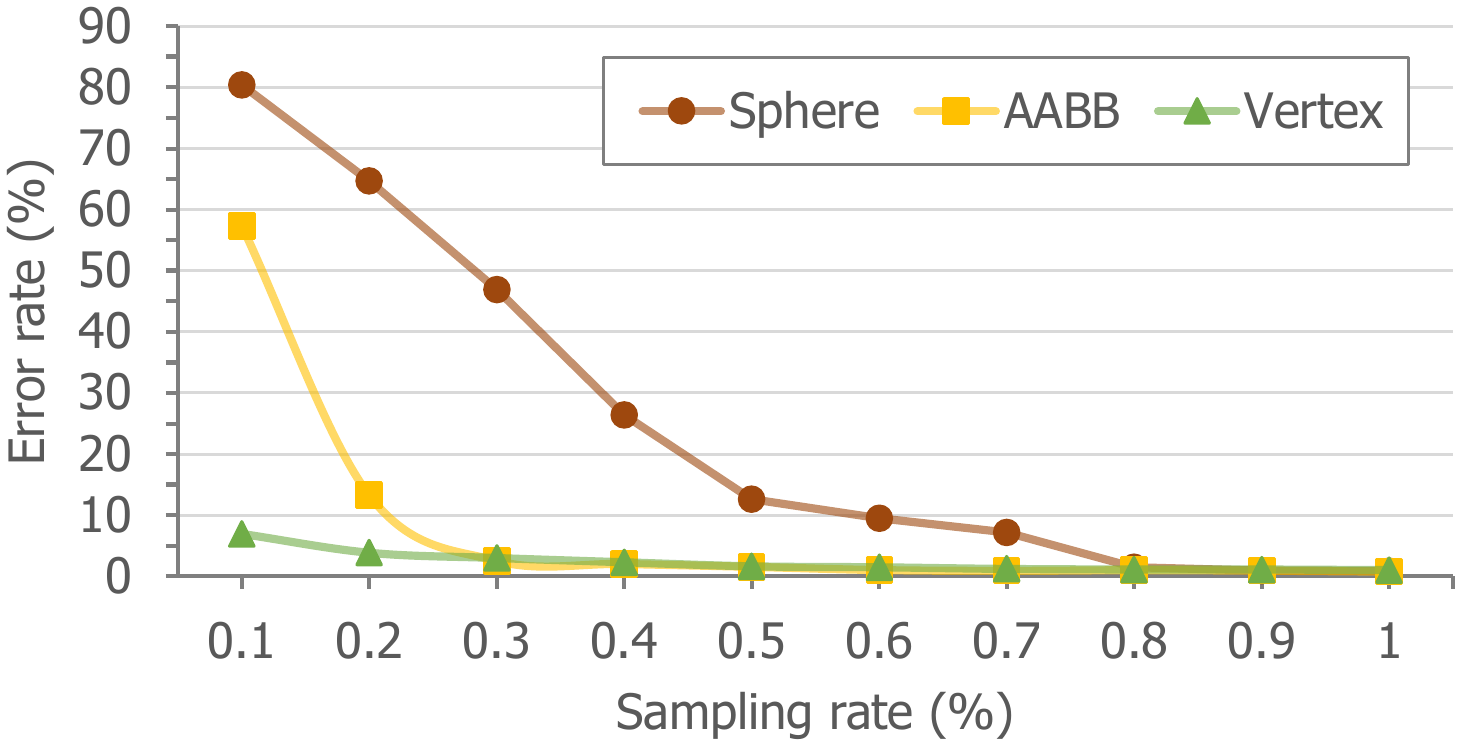}
    \caption{These graphs display changes in error rates as the sampling rate increases for three different sampling strategies in the $Lucy$ benchmark with a 0.5 overlap ratio.}
    \label{fig_sampling_method}
\end{figure}

Fig.~\ref{fig_sampling_method} shows the results for the benchmark scene of $Lucy$ with a 0.5 overlap ratio.
When the sampling rate approaches 1.0\%, all sampling strategies converge to a similar error rate (around 1\%).
Additionally, the alternative methods show slightly better accuracy in some cases when the sampling rate exceeds 1\%.
However, the vertex sampling method achieves a low error rate with smaller sampling rates (less than 0.1\%) while demonstrating the fastest convergence.

It is important to note that processing time increases linearly with the number of samples (Fig.~\ref{fig:sampling_rate}).
Therefore, while the AABB sampling method can achieve similar accuracy at a sampling rate of 0.3\%, it requires much more processing time, negating the benefit of using RT-cores.
Furthermore, we found that the alternative sampling strategies increased the processing time of RT-HDIST by about 1.7 times compared to the vertex sampling method due to the added complexity of the sampling process.


\Skip{ 
\subsection{Overlap region test}
\label{section:overlap_test}

\begin{figure}[h]
    \centering
    \begin{subfigure}[b]{0.9\linewidth}
        \centering
    \begin{tikzpicture}
    \begin{axis}[
    width=\linewidth,
    xlabel={Overlap ratio (\%)},
    ylabel={Execution time (ms)},
    legend pos=north east,
    ymajorgrids=true,
    grid style=dashed,
    ymode=log,
]
\addplot[color=blue,  mark=square] table [x=Rate, y=sh_cpu, col sep=comma] {Context/Rawdata/overlapPerf.csv};
\addplot[color=brown, mark=square] table [x=Rate, y=sh_nv, col sep=comma] {Context/Rawdata/overlapPerf.csv};
\addplot[color=orange,mark=square] table [x=Rate, y=sh_rt, col sep=comma] {Context/Rawdata/overlapPerf.csv};
\legend{$baseline$,$GPU_{naive}$,$RTPD$}
    \end{axis}
\end{tikzpicture}
\caption{Total}
\end{subfigure}
\begin{subfigure}[b]{0.9\linewidth}
        \centering
    \begin{tikzpicture}
    \begin{axis}[
    width=\linewidth,
    xlabel={Overlap ratio (\%)},
    ylabel={Execution time (ms)},
    legend pos=north east,
    ymajorgrids=true,
    grid style=dashed,
    ymode=log,
]
\addplot[color=blue,  mark=square] table [x=Rate, y=sh_cpu, col sep=comma] {Context/Rawdata/overlapPerf_PIP.csv};
\addplot[color=orange,mark=square] table [x=Rate, y=sh_rt, col sep=comma] {Context/Rawdata/overlapPerf_PIP.csv};
\legend{$baseline$,$RTPD$}
    \end{axis}
\end{tikzpicture}
\caption{PIP}
\end{subfigure}
\begin{subfigure}[b]{0.9\linewidth}
        \centering
    \begin{tikzpicture}
    \begin{axis}[
    width=\linewidth,
    xlabel={Overlap ratio (\%)},
    ylabel={Execution time (ms)},
    legend pos=north east,
    ymajorgrids=true,
    grid style=dashed,
    ymode=log,
]
\addplot[color=blue,  mark=square] table [x=Rate, y=sh_cpu, col sep=comma] {Context/Rawdata/overlapPerf_Hdist.csv};
\addplot[color=brown, mark=square] table [x=Rate, y=sh_nv, col sep=comma] {Context/Rawdata/overlapPerf_Hdist.csv};
\addplot[color=orange,mark=square] table [x=Rate, y=sh_rt, col sep=comma] {Context/Rawdata/overlapPerf_Hdist.csv};
\legend{$baseline$,$GPU_{naive}$,$RTPD$}
    \end{axis}
\end{tikzpicture}
\caption{Hausdorff}
\end{subfigure}
    \caption{The execution time of penetration depth as changes overlap ratio of $stmatthew_{High}$. $GPU_{naive}$ is brute force Hausdorff computation on GPU. We use the $PIP_{RT}$ for this test.}
    \label{fig:overlap_performance}
\end{figure}

To confirm the tendency of changes in overlap regions, we implemented and tested the special benchmark on RTX 4080. We computed the transformation to the X-axis so that the overlap region was 10, 20, 30, 40, and 50\% per each polygon.
The figure~\ref{fig:overlap_performance} presents the graph of changes in execution time according to the overlap ratio for $stmatthew_{High}$.
We found that the executive time increases as the overlap ratio. While the PIP performance is almost maintained, the Hausdorff distance is changed greatly because the Hausdorff distance is dependent on the penetration surface.
As a result, we confirm the performance of Hausdorff distance \ToCheck{4.97} times to \ToCheck{5.68} times slightly up as the overlap ratio increased than $baseline$.
This tendency demonstrates that our method is more efficient as the overlap ratio increases for the huge model.

\TODO{Time graph has too much gap. How does change to the "amount of change" graph?}

\subsection{Comparison of sampling methods}

To evaluate the impact of AABB sampling methods, we implemented both sampling methods $RTHD_{Hemisphere}$ and $RTHD_{AABB}$ and brute force Hausdorff distance method $HD_{GPU_{Naive}}$ on the GPU. Also, we set the benchmark $Lucy$ with sampling counts 10, 25, 50, 75, 100, 150, 200, and 250. 
Figure~\ref{fig:comparison_sampling_methods} presents the change of results when the sample increased.

The ray sampling time takes very little time (<=0.1\% of the total processing time) regardless of sampling count, but the total computation time linearly increases as shown in the graph. Besides the error rate decreases as the sample increases. In this benchmark, we found that the result converges in about 10,000 ray samples. In fact, $RTHD_{AABB}$ has an error rate of less than 3\% in 10000 samples. On the other hand, $RTHD_{Hemisphere}$ also converges but the error rate does not decrease more than \ToCheck{6.4}\%. This means the ray of $RTHD_{Hemisphere}$ has not been sampled in a valid direction. As a result, $RTHD_{AABB}$ achieves more accurate results within the same time at the same number of samples than $RTHD_{Hemisphere}$. This result demonstrates the efficiency of our method $RTHD_{AABB}$.
} 

\subsection{Ablation study} 

To evaluate the contribution of each component of our method to overall performance improvement, we conducted an ablation study using the $Lucy$ benchmark on an RTX 3080.
The results are shown in Table~\ref{table_ablation}.
For \revision{PPE} and HDIST, the baselines are CUDA-based implementations from $GPU_{cuda}$.
For PSG, the baseline is the CPU implementation, as we propose a GPU-based PSG algorithm in this work (Sec.~\ref{subsec:surfaceGen}).
\begin{table}[t]
\centering
\caption{Processing time (in seconds) with (\checkmark) and without our method for each component (\revision{PPE}, PSG, HDIST) on the $Lucy$ benchmark, executed on an RTX 3080.}
\label{table_ablation}
\small
\begin{tabular}{|c|ccc|ccccc|}
\hline
Type &
  \multicolumn{3}{c|}{Component} &
  \multicolumn{5}{c|}{Overlap ratio} \\ \cmidrule{2-9} 
ID &
  \revision{PPE} &
  PSG &
  HDIST &
  \multicolumn{1}{c|}{0.1} &
  \multicolumn{1}{c|}{0.3} &
  \multicolumn{1}{c|}{0.5} &
  \multicolumn{1}{c|}{0.7} &
  0.9 \\ \hline
1 &
   &
   &
   &
  \multicolumn{1}{c|}{11.00} &
  \multicolumn{1}{c|}{17.38} &
  \multicolumn{1}{c|}{26.68} &
  \multicolumn{1}{c|}{41.14} &
  46.37 \\ \hline
2 &
  \checkmark &
   &
   &
  \multicolumn{1}{c|}{10.21} &
  \multicolumn{1}{c|}{15.88} &
  \multicolumn{1}{c|}{24.99} &
  \multicolumn{1}{c|}{37.30} &
  42.06 \\ \hline
3 &
   &
  \checkmark &
   &
  \multicolumn{1}{c|}{8.20} &
  \multicolumn{1}{c|}{12.44} &
  \multicolumn{1}{c|}{19.68} &
  \multicolumn{1}{c|}{32.40} &
  36.96 \\ \hline
4 &
   &
   &
  \checkmark &
  \multicolumn{1}{c|}{7.20} &
  \multicolumn{1}{c|}{13.16} &
  \multicolumn{1}{c|}{21.24} &
  \multicolumn{1}{c|}{30.29} &
  35.16 \\ \hline
5 &
  \checkmark &
  \checkmark &
   &
  \multicolumn{1}{c|}{7.27} &
  \multicolumn{1}{c|}{10.73} &
  \multicolumn{1}{c|}{17.72} &
  \multicolumn{1}{c|}{28.36} &
  32.36 \\ \hline
6 &
  \checkmark &
   &
  \checkmark &
  \multicolumn{1}{c|}{5.29} &
  \multicolumn{1}{c|}{9.56} &
  \multicolumn{1}{c|}{14.45} &
  \multicolumn{1}{c|}{18.59} &
  20.53 \\ \hline
7 &
   &
  \checkmark &
  \checkmark &
  \multicolumn{1}{c|}{4.33} &
  \multicolumn{1}{c|}{8.48} &
  \multicolumn{1}{c|}{14.54} &
  \multicolumn{1}{c|}{21.62} &
  26.12 \\ \hline
8 & \checkmark & \checkmark & \checkmark & \multicolumn{1}{c|}{2.39} & \multicolumn{1}{c|}{5.29} & \multicolumn{1}{c|}{9.27} & \multicolumn{1}{c|}{13.27} & 15.52 \\ \hline
\end{tabular}
\end{table}

As shown in rows 2 to 4 of Table~\ref{table_ablation}, all components contribute to reducing the total processing time.
Among the three, the RT-based HDIST algorithm has the greatest impact, reducing processing time by about 26\% on average.
The GPU-based PSG algorithm also plays a critical role, achieving the highest relative performance improvement compared to the baseline, despite having the smallest workload (Table~\ref{table:breakdown}), with an average reduction of 24\%.
While the RT-based PPE contributes the least to overall performance, it still consistently reduces total processing time by about 8\% on average across all tested cases.

As we incorporate more components from our method, the performance improves significantly compared to using just one component. Specifically, types 5, 6, and 7 reduce the total processing time by approximately 33\%, 51\%, and 50\% on average, respectively.
Finally, the full version of our method (type 8, $RTPD$) achieves the best performance, reducing the processing time by about 69\% on average.

Although $RTPD$ is an approximate penetration depth algorithm due to its sample-based HDIST computation, our method can be applied only to the PPE and PSG stages while using the CUDA-based HDIST method if an application requires exact results.
This approach is demonstrated in type 5 of Table~\ref{table_ablation}, which still improves performance by 1.50 times on average over the baseline (type 1).
These results demonstrate that the components of our method can be selectively applied to various applications based on specific requirements, leading to improved performance.

\Skip{ 
Table~\ref{table:Ablation study} presents the result of implementation that applies our methods for each step.
Specifically, our RT-based PIP methods have a lot more benefits than $baseline$'s PIP (almost \ToCheck{x10.28} times faster. It can compute compared with test type 6 and RTPD).
In a similar manner, we can compute the benefit of the penetration surface generation method and RT-based Hausdorff distance, they achieve \ToCheck{x2.24} and \ToCheck{x8.91} times faster.
On the other hand, the accuracy (error rate) of $baseline$ is better than our methods (See $baseline$ and test types 1,2,4).
Instead, we save the execution times at the expense of accuracy (We get times up \ToCheck{x18.57} times than $baseline$ and lose the accuracy by \ToCheck{1.49}\%). 
These results demonstrate that our RT-based method is more efficient than $baseline$ methods.
} 


\section{Conclusion}

We introduced RTPD, a novel algorithm for calculating penetration depth using hardware-accelerated ray-tracing cores (RT-cores).
Our approach leverages the specialized capabilities of RT-cores to efficiently perform penetration surface extraction and Hausdorff distance calculations.
We also presented a GPU-based algorithm for generating penetration surfaces, ensuring that RTPD operates entirely on GPU platforms.
Through extensive testing on various generations of RTX GPUs and benchmark scenes, our algorithm demonstrated significant performance improvements, outperforming a state-of-the-art penetration depth method and conventional GPU implementations by up to 37.66 times and 5.33 times, respectively.

These results highlight the potential of RT-cores beyond their traditional rendering applications, suggesting broad applicability in diverse computational tasks, including simulations, the metaverse, and robotics. 
The efficiency and scalability of RTPD make it a promising solution for applications that require precise and rapid penetration depth calculations.

\textbf{Limitations and Future Work:}
Since our RT-based Hausdorff distance computation uses a sampling method, the results may not match exact penetration depth values.
Future work will focus on finding more efficient sampling strategies to reduce errors further.
We also aim to develop RT-based penetration depth algorithms that guarantee exact results.
Additionally, we plan to explore the application of RT-core-based algorithms in other domains, expanding their use in various computational tasks.

\bibliographystyle{abbrv-doi-hyperref}
\bibliography{ref}

\end{document}